\documentclass[twocolumn]{aastex631}

\usepackage{amsmath}	% Advanced maths commands
\usepackage{amssymb}	% Extra maths symbols

\usepackage{CJK}

\newcommand{\renc}{r_{\text{enc}}} % radius of enclosing sphere 
\newcommand{\rvol}{r_{\text{vol}}} % radius of sphere of equivalent volume
\newcommand{\rpja}{r_{\text{pja}}} % radius of circle of equivalent projected area

\newcommand{\xenc}{x_{\text{enc}}}
\newcommand{\xvol}{x_{\text{vol}}}

\newcommand{\Cext}{C_{\text{ext}}}
\newcommand{\Cabs}{C_{\text{abs}}}
\newcommand{\Csca}{C_{\text{sca}}}

\newcommand{\kabs}{\kappa_{\text{abs}}}
\newcommand{\ksca}{\kappa_{\text{sca}}}

\newcommand{\Qextenc}{Q_{\text{ext}}^{\text{enc}}}
\newcommand{\Qabsenc}{Q_{\text{abs}}^{\text{enc}}}

\newcommand{\Qextvol}{Q_{\text{ext}}^{\text{vol}}}
\newcommand{\Qabsvol}{Q_{\text{abs}}^{\text{vol}}}

\newcommand{\Qextpja}{Q_{\text{ext}}^{\text{pja}}}
\newcommand{\Qabspja}{Q_{\text{abs}}^{\text{pja}}}

\newcommand{\vect}[1]{\boldsymbol{#1}}
\newcommand{\matr}[1]{\mathbf{#1}}

\newcommand{\nlat}{n_{\text{lat}}}

\newcommand{\expnumber}[2]{{#1}\mathrm{e}{#2}}

\begin{document}

\title{
\texttt{glitterin}: 
Towards Replacing the Role of Lorenz-Mie Theory in Astronomy Using \\ Neural Networks Trained on Light Scattering of Irregularly Shaped Grains
}

\author[0000-0001-7233-4171]{Zhe-Yu Daniel Lin (\begin{CJK*}{UTF8}{bkai}林哲宇\end{CJK*})}
\thanks{Jansky Fellow of the National Radio Astronomy Observatory}
\affiliation{Earth and Planets Laboratory, Carnegie Science, 5241 Broad Branch Rd NW, 
Washington, DC 20015, USA
}
\affiliation{National Radio Astronomy Observatory, 520 Edgemont Road, Charlottesville, VA 22903, USA
}
\email{zlin@carnegiescience.edu}
\email{dlin@nrao.edu}

\author[0000-0001-6654-7859]{Alycia J. Weinberger}
\affiliation{Earth and Planets Laboratory, Carnegie Science, 
5241 Broad Branch Rd NW, 
Washington, DC 20015, USA
}

\author[0000-0001-9994-923X]{Evgenij Zubko}
%\affiliation{Planetary Atmospheres Group, Institute for Basic Science (IBS), 55 Expo-ro, Yuseong-gu, Daejeon 34126, South Korea}
\affiliation{Department of Physics, P.O. Box 64, University of Helsinki FI-00014, Finland}

\author[0000-0001-7824-5372]{Jessica A. Arnold}
\affiliation{Army Research Laboratory, 2800 Powder Mill Rd, Adelphi, MD 20783, USA}

\author[0000-0002-4177-7364]{Gorden Videen}
\affiliation{
Space Science Institute, 4750 Walnut Street, Boulder Suite 205, 
CO 80301, USA
}
\affiliation{
Kyung Hee University, Yongin, South Korea
}
\affiliation{
Department of Atmospheric Sciences, Texas A\&M University, College Station, TX, USA
}

%% Note that the \and command from previous versions of AASTeX is now
%% depreciated in this version as it is no longer necessary. AASTeX 
%% automatically takes care of all commas and "and"s between authors names.

%% AASTeX 6.31 has the new \collaboration and \nocollaboration commands to
%% provide the collaboration status of a group of authors. These commands 
%% can be used either before or after the list of corresponding authors. The
%% argument for \collaboration is the collaboration identifier. Authors are
%% encouraged to surround collaboration identifiers with ()s. The 
%% \nocollaboration command takes no argument and exists to indicate that
%% the nearby authors are not part of surrounding collaborations.

%% Mark off the abstract in the ``abstract'' environment. 

\begin{abstract}

Light scattering by dust particles is often modeled assuming the dust is spherical for numerical simplicity and speed. However, real dust particles have highly irregular morphologies that significantly affect their scattering properties. We have developed \texttt{glitterin}, a neural network trained to predict light scattering from irregularly shaped dust grains, offering a computationally efficient alternative to Lorenz-Mie theory.
We computed scattering properties using the Discrete Dipole Approximation code ADDA for irregularly shaped particles across size parameters $x$ from 0.1 to 65, covering a range in complex refractive index $m$ that includes astrosilicates, pyroxene, enstatite, water-ice, etc. The neural network operates at millisecond timescales while maintaining superior accuracy compared to linear interpolation.
Irregular grains exhibit $x$-dependent deviations from spherical predictions. At small $x$, cross-sections approach volume-equivalent spheres for low $m$. At large $x$, irregular grains show enhanced cross-sections due to greater geometric extension. Increasing $m$ also enhances the absorption cross-section relative to the volume-equivalent spheres.
This differential $x$ and $m$ dependence creates mid-IR solid-state features distinct from predictions from spherical grains. 
Validation against laboratory measurements of forsterite and hematite demonstrates that our neural network captures both qualitative and quantitative behaviors more accurately than spherical models. Millimeter-wavelength applications reveal that spherical grains produce opposite polarization signatures compared to irregular grains, potentially relaxing stringent $\sim100$~$\mu$m grain size constraints in protoplanetary disks. 
\texttt{glitterin} is publicly available and alleviates the computational barriers to incorporating emission and scattering of realistic grain morphologies in the dust inference and radiative transfer simulations for debris disks and other astronomical environments.

\end{abstract}

%% Keywords should appear after the \end{abstract} command. 
%% The AAS Journals now uses Unified Astronomy Thesaurus concepts:
%% https://astrothesaurus.org
%% You will be asked to selected these concepts during the submission process
%% but this old "keyword" functionality is maintained in case authors want
%% to include these concepts in their preprints.

%\keywords{Circumstellar disks (235) --- Debris disks (363) --- Dust continuum emission (412) --- Interstellar dust (836) ---Neural Networks}

%% From the front matter, we move on to the body of the paper.
%% Sections are demarcated by \section and \subsection, respectively.
%% Observe the use of the LaTeX \label
%% command after the \subsection to give a symbolic KEY to the
%% subsection for cross-referencing in a \ref command.
%% You can use LaTeX's \ref and \label commands to keep track of
%% cross-references to sections, equations, tables, and figures.
%% That way, if you change the order of any elements, LaTeX will
%% automatically renumber them.
%%
%% We recommend that authors also use the natbib \citep
%% and \citet commands to identify citations.  The citations are
%% tied to the reference list via symbolic KEYs. The KEY corresponds
%% to the KEY in the \bibitem in the reference list below. 

\section{Introduction} \label{sec:intro}

Dust grains are the main constituent of debris disks, which are replenished by constant collisions of planetesimal or planetary bodies \citep[e.g.][]{Hughes2018ARA&A..56..541H}. 
Telescopes, such as the HST, VLT/SPHERE, Magellan-Clay Telescope, and Gemini/GPI, etc, observing in the visible and near infrared wavelengths, detect the stellar photons scattered off the dust \citep[e.g.][]{Rodigas2015ApJ...798...96R, Xie2022A&A...666A..32X}. Several studies have revealed prominent forward-scattering behavior, which is indicative of grains much larger than the wavelength of light \citep[e.g.][]{Close1998ApJ...499..883C, Greenberg1981A&A....93...35G, Kalas2013ApJ...775...56K, Milli2017A&A...599A.108M, Milli2019A&A...626A..54M, Chen2020ApJ...898...55C, Olofsson2024A&A...688A..42O}. 
Resolved images of polarization also have been obtained, which are expected to constrain the dust size distribution and composition \citep[e.g.][]{Milli2019A&A...626A..54M, Arriaga2020AJ....160...79A, Esposito2020AJ....160...24E, Olofsson2024A&A...688A..42O, Hom2025AJ....170...46H}.
Spitzer and JWST enabled access to the thermal emission in the mid-infrared from warm grains in the debris disk and revealed abundant solid-state features from various species, like silicates
\citep[e.g.][]{Olofsson2012A&A...542A..90O, Mittal2015ApJ...798...87M, Chai2024ApJ...976..167C, Gaspar2023NatAs...7..790G, Su2024ApJ...977..277S, Xie2025Natur.641..608X, Chen2024ApJ...973..139C}. 
The Atacama Large Millimeter/submillimeter Array (ALMA) offers a millimeter wavelength window that can probe the thermal emission from colder and larger $\sim$mm grains \citep[e.g.][]{MacGregor2017ApJ...842....8M, Matra2020ApJ...898..146M, Terrill2023MNRAS.524.1229T}.

Proper inferences of the dust grain properties, such as the size distribution, spatial distribution, and composition, can allow constraints on the dynamical processes in disks or planet-disk interaction \citep[e.g.][]{Wyatt2008ARA&A..46..339W, Matthews2014prpl.conf..521M}.  
In order to understand how light is emitted or scattered from these large grains, a common approach is to assume spherical grains, which permits exact solutions \citep{Lorenz1890, Mie1908AnP...330..377M}. 
However, numerous studies have established that light scattering of spherical grains is very different from that of randomly oriented irregular grains \citep[e.g.][]{Munoz2011JQSRT.112.1646M, 
Munoz2021ApJS..256...17M, Martikainen2025MNRAS.537.1489M}. 
Indeed, assuming spheres as a grain model in debris disks has resulted in difficulties in convergence, including unrealistic results when fitting multiwavelength data or spectra \citep[e.g.][]{Rodigas2015ApJ...798...96R, Lu2022ApJ...933...54L} or fitting scattering phase functions or polarization \citep[e.g.][]{Milli2019A&A...626A..54M, Duchene2020AJ....159..251D, Arriaga2020AJ....160...79A}. 
Unless the grains are actually nearly spherical, applying Lorenz-Mie theory can give misleading results. 

Despite this widely known issue, moving beyond spheres to consider irregular grains has proven difficult.
Numerical techniques, like the Discrete Dipolar Approximation (DDA; \citealt{Purcell1973ApJ...186..705P, Draine1994JOSAA..11.1491D}) and the T-matrix method \citep{Mishchenko1996JQSRT..55..535M}, can solve the light scattering of a grain of any shape, but the high computational demand has limited broad parameter space searches that are necessary to infer dust properties and its uncertainty. 
Other simplifications exist. 
The Continuous Distribution of Ellipsoids \citep{Bohren1983asls.book.....B}) or the Distribution of Hollow Spheres method \citep{Min2003A&A...404...35M, Min2005A&A...432..909M} utilizes populations of simple geometric shapes (ellipsoids of varied shapes or hollow spheres of different sizes) where exact solutions exist, and aim to approximate the population average of irregular grains. 
The Effective Medium Theory applies Lorenz-Mie theory of a modified refractive index that mixes vacuum according to a mixing rule \citep{Cuzzi2014ApJS..210...21C, Kataoka2014A&A...568A..42K}. 
These methods offer simplicity for large parameter space searches, but at the cost of systematic differences with irregular grains \citep[e.g.][]{Shen2008ApJ...689..260S, Tazaki2016ApJ...823...70T, Kirchschlager2013A&A...552A..54K,
Min2016A&A...585A..13M, Arnold2019AJ....157..157A}.
Empirical phase functions, such as a single or multiple Henyey-Greenstein phase functions \citep{Henyey1941ApJ....93...70H} offer the flexibility to provide phase functions that observations require; however, it is not possible to infer dust properties from such functions as they are not rooted in the Maxwell equations and these phase functions are not expressed as physical properties of the particles. 
Laboratory measurements are another valuable resource, but costly and limited in the parameter space \citep[e.g.][]{TobonValencia2022A&A...666A..68T, TobonValencia2024A&A...688A..70T, Munoz2025JQSRT.33109252M}. 
Their primary benefit lies in testing grain models rather than in producing quantifiable dust properties from observations. 
%\red{(Oloffson+2012) For example, samples come in a size distribution and information about the contribution from different grain sizes is difficult to obtain. }
Lookup tables of light scattering of irregular grains allows accurate and fast analysis when combined with interpolation, but these are often predetermined at specific sizes, wavelengths, or a series of compositions that limits generalizability \citep[e.g.][]{Arnold2019AJ....157..157A, Arnold2022ApJ...930..123A, Tazaki2023ApJ...944L..43T, Jang2024A&A...691A.148J}.

With the plethora of observations, there is a strong demand in the community for dust models that are both accurate and fast to allow proper inferences of dust across a wide variety of potential compositions. 
Similar issues exist for other astronomical environments where dust grains are likely not spherical,  
including protoplanetary disks \citep[e.g.][]{Lin2023MNRAS.520.1210L}, 
protostellar envelopes \citep[e.g.][]{Slavicinska2025ApJ...986L..19S}, 
or exoplanet atmospheres \citep[e.g.][]{Gao2021JGRE..12606655G, Vahidinia2024PASP..136h4404V}, 
etc.  
While new methods of efficient and accurate light-scattering calculations are constantly being developed, the community needs an adequate placeholder before these methods mature.

In the context of cosmic dust, the detailed physics of how a single particle of a specific shape scatters light is often not important, but it is the connection to grain sizes and composition as an ensemble that provide the interesting aspects of dust evolution in debris disks and other environments. 
%Existing light-scattering models have already been treated as a black box to a certain degree, which motivates us to search for an efficient emulator. 
In this paper, we experiment with the idea that we can train a neural network to learn the emission and scattering properties of an ensemble of irregular grains calculated by DDA. 
There have already been successful examples in the atmospheric sciences. 
\cite{Chen2022GeoRL..4997548C} trained a neural network on ensembles of spheroids, while \cite{Wang2023JGRD..12839568W} trained a neural network on ensembles of fractal black carbon particles. 
The problem specific to astronomy lies in the broad possibility of different dust species. 

%a database is necessary for exploration of a variety of compositions, instead of calculating each composition for a certain grain size and wavelength. the coverage in n,k can overlap. 

We introduce \texttt{glitterin} (Dust Grain Light-Scattering Neural Network), a neural network trained on light-scattering of irregularly shaped grains for a broad range of refractive index. 
The paper is organized as follows.
Sec.~\ref{sec:database} explains our scattering simulation setup, including creating irregular grains and using a publicly available DDA code. 
Sec.~\ref{sec:neural_network} describes how the neural network was trained and its evaluation. 
We discuss and demonstrate various applications in Sec.~\ref{sec:discussion} and conclude in Sec.~\ref{sec:conclusion}. 

\section{Scattering Simulations} \label{sec:database}

In this section, we describe the methodology in calculating the emission and light-scattering properties of an ensemble of irregularly shaped grains using DDA. 
% With the expected size distributions in debris disks, emission or scattered light from a population of grains are mostly dominated by grains with sizes that are comparable to the observing wavelength \citep{Hughes2018ARA&A..56..541H}. DDA is particularly useful in this regime. 
The polarized radiation transfer equation at a given frequency $\nu$ is
\begin{align}
    \frac{d}{ds} \vect{I}(s)
    = 
    - n_{d} \Cext \vect{I}(s)
    + n_{d} \Cabs \vect{B} 
    + n_{d} \oint \matr{Z}(\Omega') \vect{I}_{\text{inc}}
    d \Omega'
\end{align}
where $\vect{I}=(I,Q,U,V)^{T}$ is the Stokes parameter, $\vect{B}=(B_{\nu}(T),0,0,0)^{T}$ represents the non-polarized, blackbody thermal emission of the dust with a given temperature $T$, $\vect{I}_{\text{inc}}$ is the Stokes parameters of the incoming radiation, $s$ is the path length along the line of sight, $n_{d}$ is the number density of the dust (in counts per unit volume), and $\matr{Z}$ is the $4 \times 4$ scattering matrix that depends on the angle between the incoming and outgoing radiation \citep[e.g.][]{vandeHulst1957lssp.book.....V}. 
The last term integrates over all steradian $\Omega'$ of the incoming radiation and scatters to the line of sight. $\Cext$ and $\Cabs$ are the extinction and absorption cross-sections, respectively. The wavelength of light $\lambda$ and the properties of the dust grains, including its size, composition, and shape, determines $\Cext$, $\Cabs$, and $\matr{Z}$, which dictates the observed radiation.

By invoking random orientation, $\matr{Z}$ only depends on the scattering angle $\theta$, which is the angle between the direction of propagation of the incident and outgoing ray. In addition, only 6 elements of $\matr{Z}$ are independent and non-zero, namely: 
\begin{align}
    \matr{Z} = 
    \begin{pmatrix}
        Z_{11} & Z_{12} & 0 & 0 \\
        Z_{12} & Z_{22} & 0 & 0 \\
        0 & 0 & Z_{33} & Z_{34} \\
        0 & 0 & - Z_{34} & Z_{44}
    \end{pmatrix}
\end{align}
\citep{vandeHulst1957lssp.book.....V}. 
The scattering cross section is $\Csca = \Cext - \Cabs$ from conservation of energy and is related to $Z_{11}$ by
\begin{align}
    \Csca = 2 \pi \int_{-1}^{1} Z_{11} (\mu) d \mu
\end{align}
where $\mu \equiv \cos \theta$. It is convenient to define the single-scattering albedo and its complementary value as
\begin{align} \label{eq:albedo_ems}
    \omega \equiv \frac{ \Csca }{ \Cext } \text{ , } \epsilon \equiv 1 - \omega \text{ .}
\end{align}
In addition, it is convenient to define the following normalized quantities:
\begin{align}
    N_{11} &\equiv \frac{2\pi Z_{11}}{\Csca}, 
    &N_{12} &\equiv - \frac{Z_{12}}{Z_{11}}, 
    & N_{22} &\equiv \frac{Z_{22}}{Z_{11}}, \nonumber \\
    N_{33} &\equiv \frac{Z_{33}}{Z_{11}}, 
    &N_{34} &\equiv \frac{Z_{34}}{Z_{11}}, 
    & N_{44} &\equiv \frac{Z_{44}}{Z_{11}}
\end{align}
where $N_{12}$ is called the Degree of Linear Polarization (DLP). $N_{11}$ is also called the phase function. 

\subsection{Particle Setup}

The specific shape of grains in debris disks is poorly understood. As a result, there are several algorithms to produce irregular grains, including Gaussian Random Field particles \citep[e.g.][]{Muinonen1996JQSRT..55..577M, Min2007A&A...462..667M} or Ballistic Agglomeration that incorporates physical growth rules \citep[e.g.][]{Shen2008ApJ...689..260S}. 
Furthermore, due to the similarity of the light-scattering properties of highly irregularly shaped particles despite having different morphological properties \citep[e.g.,][]{Zubko2013MNRAS.430.1118Z}, the precise prescription for the shape appears unimportant.
We adopt an algorithm based on the agglomerated debris particle \citep{Zubko2009JQSRT.110.1741Z}. This model has shown success in replicating the laboratory measurements of the scattering matrix of irregular grains, including the phase function, degree of linear polarization, and circular polarization \citep{Zubko2013JQSRT.131..175Z, Zubko2024JQSRT.32309053Z}. Most significantly, it is able to do this at multiple wavelengths simultaneously. In addition, since it is a simple nearest-neighbor search at its core, it is easily implementable with commonly available software (e.g., \texttt{scipy}) and capable of generating as many realizations as necessary with little computational cost. 
The term ``irregular" grains implies a lack of any repetition in the structure and is guaranteed by the algorithm. 
A comparison of various irregular grain prescriptions is beyond the scope of this paper, but the analysis demonstrated here can be easily extended to other particles.

\begin{figure}
    \centering
    \includegraphics[width=\columnwidth]{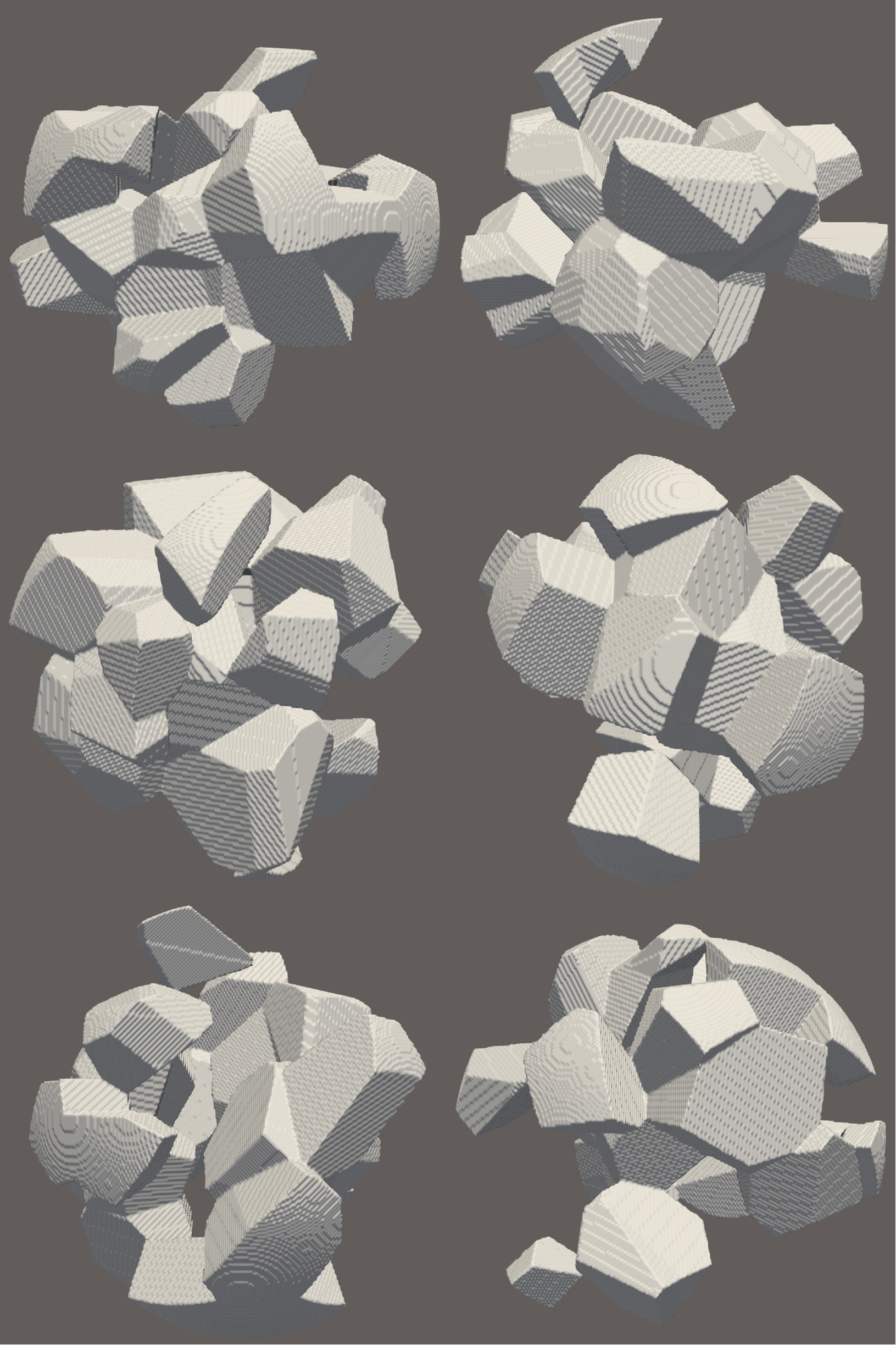}
    \caption{
        Visualization of six different irregular particles with $\nlat=256$ as a demonstration.
    }
    \label{fig:visualization}
\end{figure}

The creation of a particle is as follows. Consider an imaginary sphere of radius $r_{\text{enc}}$ in spherical coordinates $(r_{p},\theta_{p},\phi_{p})$. We randomly select 21 points within the sphere to serve as material seeds and 20 points within the sphere as vacuum seeds. The Cartesian coordinates ($x_{p},y_{p},z_{p}$) of each seed is uniformly sampled between $(-\renc,\renc)$, while rejecting those that do not satisfy $x_{p}^{2}+y_{p}^{2}+z_{p}^{2} \leq \renc^{2}$. 
Furthermore, we add 100 vacuum seeds that randomly sample the surface of the imaginary sphere. The polar and azimuthal angle of each seed is $\xi = 2 \cos \theta_{p} -1$ and $\phi_{p} = 2 \pi \xi$ where $\xi$ is a random number from $0$ to $1$ (one random number for each angle). With the 21 material seeds and 120 vacuum seeds, we can run a nearest neighbor search for a lattice of cubical dipoles of any size neccessary for DDA. We denote the number of dipoles along a dimension as $\nlat$. 
Fig.~\ref{fig:visualization} shows the visualization of a few examples (see also, e.g., \citealt{Arnold2019AJ....157..157A}).

There are two key structural parameters that characterize irregularly shaped grains. 
The volume filling factor $f$ is defined as the ratio between the volume of material and the volume of the enclosing sphere. 
The volume of material for an irregularly shaped grain is simply the number of material dipoles times the volume of each cubical dipole. 
The volume of the enclosing sphere is $4 \pi \renc^{3} / 3$. We can relate the radius of a sphere of equivalent volume $\rvol$ to $\renc$ by $f$ through
\begin{align}
    \rvol^{3} \equiv f \renc^{3} \text{.}
\end{align}
In the geometric-optics regime, $\Cext$ is twice the projected geometric area of the particle from the extinction paradox \citep[e.g.][]{vandeHulst1957lssp.book.....V}. Thus, it is also convenient to measure the projected area factor $g$ by the ratio of the projected geometric area to that of the enclosing sphere ($\pi \renc^{2}$). We define a circle with a radius that makes the area equal to the projected area by 
\begin{align}
    \rpja^{2} = g \renc^{2} \text{.}
\end{align}
We adopt $f=0.236$ and $g=0.611$ based on \cite{Zubko2024JQSRT.32309053Z}.

Because of the different radii, $\rvol$, $\rpja$, and $\renc$, there can be 3 different ways to normalize each of the 3 different cross-sections (a total of 9 different efficiency factors). 
For clarity, we denote these by
\begin{align} \label{eq:C_eq_Q_pi_r2}
    C_{i} \equiv Q_{i}^{j} \pi r_{j}^{2}
\end{align}
where $i$ can be ``ext," ``abs," or ``sca" for extinction, absorption, and scattering, respectively, and $j$ can be ``vol," ``pja," or ``enc" for the $\rvol$, $\rpja$, or $\renc$, respectively. 
For example, $\Qextenc$ is $\Cext$ normalized by $\pi \renc^{2}$ and $\Qabspja$ is $\Cabs$ normalized by $\pi \rpja^{2}$. 
Note that for spheres, $\rvol=\rpja=\renc$, so there is no distinction between how each cross-section is normalized. 
The three different radii also means there are three different size parameters
\begin{align}
    x_{i} \equiv \frac{ 2 \pi r_{i} }{ \lambda }
\end{align}
where $i$ can be ``vol," ``pja,", or ``enc."

\subsection{Discrete Dipole Approximation} \label{sec:dda}

We compute the optical properties of the dust particle using the ADDA code \citep{Yurkin2007JQSRT.106..558Y, Yurkin2011}.\footnote{\url{https://github.com/adda-team/adda}}
We use the filtered-coupled dipole (FCD) method which allows calculations for a broad range of refractive index \citep{Piller1998ITAP...46.1126P, Yurkin2010PhRvE..82c6703Y}.

In the DDA, there are two fundamental constraints between the dipole size and lattice size. 
Assuming cubical dipoles, we define a parameter
\begin{align} \label{eq:mkd_def}
    q \equiv \frac{2\pi |m| d}{ \lambda }
\end{align}
where $d$ is the length of the cubical dipole and $m \equiv n + i k$ is the complex refractive index. 
The DDA requires that the dipole size must be much smaller than the wavelength inside the material ($\lambda/|m|$), which means $q$ needs to be less than some threshold ($q \leq q_{c}$). The exact value of $q_{c}$ is unclear, but it is of order unity $q_{c} \sim 1$ \citep{Draine1994JOSAA..11.1491D}. For irregular grains, $q_{c}$ can be even slightly larger than 1 (up to $\sim 1.3$) and still maintain similar population-averaged scattering properties compared to smaller $q_{c}$ cases \citep{Zubko2010ApOpt..49.1267Z, Zubko2024JQSRT.31408854Z}. 
For this paper, we take $q_{c}=1$. 
The second constraint is in the number of dipoles required to approximate the shape of the particle. 
The number of dipoles along a dimension across the particle is related to the radius of the enclosing sphere: 
\begin{align} \label{eq:nlat_def}
    \nlat d \equiv 2 \renc \text{ .}
\end{align}
We impose a minimum number of dipoles through $\nlat \geq n_{\text{min}}=32$. 
Lastly, as part of the input parameter to ADDA, the dipole-per-lambda parameter is
\begin{align} \label{eq:dpl_def}
    \text{dpl} \equiv \frac{\lambda}{d},
\end{align}
which defines the size of the dipole with respect to the wavelength of the vacuum. 
When using FCD, dpl should be greater than 2 and thus, we consider $\text{dpl} \geq d_{\text{min}}$ and take $d_{\text{min}}=4$. This is rarely an issue in our parameter space, but it can happen when $n$ is small and when the size parameter is large. 

Two particles with the same shape and refractive index, but with different sizes will have the same optical properties as long as the size parameters are equal. 
From this aspect, it is most efficient to produce simulations in ($\xenc, n, k$) space instead of ($\renc, \lambda, n, k$), effectively reducing 1 dimension.
With Eqs.~\ref{eq:mkd_def}, \ref{eq:nlat_def}, and \ref{eq:dpl_def}, we can write an alternative set of three equations:
\begin{subequations}
\begin{align}
    q &= \frac{2 |m| \xenc }{ \nlat} \\
    q &= \frac{ 2 \pi |m| } { \text{dpl} } \\
    \frac{ \nlat }{ \text{dpl} } &= \frac{\xenc}{\pi },
\end{align}
\end{subequations}
which no longer depends on $\lambda$ and $\renc$ explicitly and only on $\xenc$ and $m$ which are the actual input parameters we care about. 

We wish to determine the maximum acceptable $q$ that satisfies all three inequalities, which means
\begin{align}
    q_{\text{max}} = \text{min} \bigg(q_{c}, \frac{2 |m| \xenc }{ n_{\text{min}}},  \frac{ 2 \pi |m| } { \text{dpl}_{\text{min}} } \bigg) \text{ .}
\end{align}
Since $\nlat$ has to be an integer when constructing the lattice of dipoles, we find the next largest $q$ such that $q < q_{\text{max}}$ and makes $\nlat$ an integer. 
%The wavelength to run ADDA is set to the default $\lambda=2\pi$~$\mu$m though the specific value of $\lambda$ here does not matter, since we are only concerned with the size parameter. 

In each ADDA simulation, we use the default convergence threshold of the iterative solver at $10^{-5}$. 
The scattering angles $\theta$ are chosen on a discrete grid defined on a Gaussian quadrature. 
Large grains exhibit strong forward scattering and finer structure in the phase function, which means we require finer spacing at small $\theta$. 
Thus, depending on the maximum $\xenc$, the $\theta$ grid can be different for each dataset. 
We also include forward ($\theta=0^{\circ}$) and backscattering ($\theta=180^{\circ}$) points. 
%The cross sections $\Cext$ and $\Cabs$ is an average of two initial polarization directions. 
We average $Z_{ij}$ over 16 scattering planes. 

To calculate the ensemble average of the population of agglomerated debris particles, we use 500 instances for any combination of ($\xenc$, $n$, $k$) to obtain the population average of $\Cext$, $\Cabs$, and $Z_{ij}$. Each instance is a different grain structure. 
Aside from the average, we also calculate the uncertainty of the average cross-sectional quantities, $\delta \Cext$, $\delta \Cabs$, and $\delta Z_{ij}$, by bootstrapping (through \texttt{bootstrap} from \texttt{scipy} using \texttt{n\_resamples}=1000). 
The uncertainties of normalized quantities, like $\delta \omega$, $\delta \epsilon$, and $\delta N_{ij}$, are also estimated by bootstrapping, since there is significant correlation in the cross-sectional quantities. 
The chosen number of instances generally allows the $\delta N_{12}$ to be $\sim 1\%$ consistent with previous calculations \citep{Zubko2013JQSRT.131..175Z}. 
The quantities $Z_{13}$, $Z_{14}$, $Z_{23}$, $Z_{24}$, $Z_{31}$, $Z_{32}$, $Z_{41}$, and $Z_{42}$ are 0 as expected from an ensemble of randomly oriented grains \citep{vandeHulst1957lssp.book.....V}. 

Throughout the paper, we often make comparison to the spherical case by using the \texttt{bhmie} code from the \texttt{dsharp\_opac} package for the Lorenz-Mie calculations \citep{Birnstiel2018ApJ...869L..45B}. 

\subsection{Refractive-Index Coverage} \label{sec:composition}

% astronomical silicate \citep{Draine2003ARA&A..41..241D}. 
% amorphous pyroxene \citep{Jaeger1994A&A...292..641J}
% organic refractory \citep{Li1997A&A...323..566L}

Since the complex refractive index for a given dust species can vary significantly across wavelength, the creation of the database does not assume any particular species, but samples the ($n,k$)-space to maximize generalizability. 

We use Sobol sampling (implemented in \texttt{scipy}) for generating the distribution of $\xenc$, $n$, and $k$ \citep{Sobol196786}. The sampling generates more uniform quasi-random coverage in the parameter space than random sampling \citep{Kaeufer2023A&A...672A..30K}. 
%and should perform better than gridded sampling for evaluating the optical properties across the entire parameter space \citep{Bergstra_2012}. 
Both $\xenc$ and $k$ are uniformly sampled in log space, while $n$ is uniformly sampled linearly. 

We impose boundaries in the parameter space due to limitation in computational time and to focus on species of interest. 
We are mostly interested in the red optical to mid-IR range ($0.7 \sim 20$~$\mu$m) and we consider the following: crystalline olivine (Mg$_{1.72}$Fe$_{0.21}$SiO$_{4}$) \citep{Zeidler2015ApJ...798..125Z}, 
enstatite (MgSiO$_{3}$) \citep{Jaeger1998A&A...339..904J}, 
amorphous carbon \citep{Zubko1996MNRAS.282.1321Z}, 
water ice \citep{Warren2008JGRD..11314220W}, and troilite \citep{Henning1996A&A...311..291H}. 
In addition, we also consider the ``DSHARP mixture'' from the Disk Substructures at High Angular Resolution Project (DSHARP) in protoplanetary disks \citep{Birnstiel2018ApJ...869L..45B}, which assumes a mixture of water-ice \citep{Warren2008JGRD..11314220W}, astronomical silicates \citep{Draine2003ARA&A..41..241D}, troilite and refractory organics \citep{Henning1996A&A...311..291H}. 
The range covers the DSHARP prescription up to $\lambda\sim9$~mm meaning the scattering data can be applied to micron-size grains in the mid-IR (Sec.~\ref{sec:10_micron_complex}) or millimeter-size grains in the (sub)millimeter wavelength regime (Sec.~\ref{sec:millimeter_wavelength}).

\begin{figure*}
    \centering
    \includegraphics[width=\linewidth]{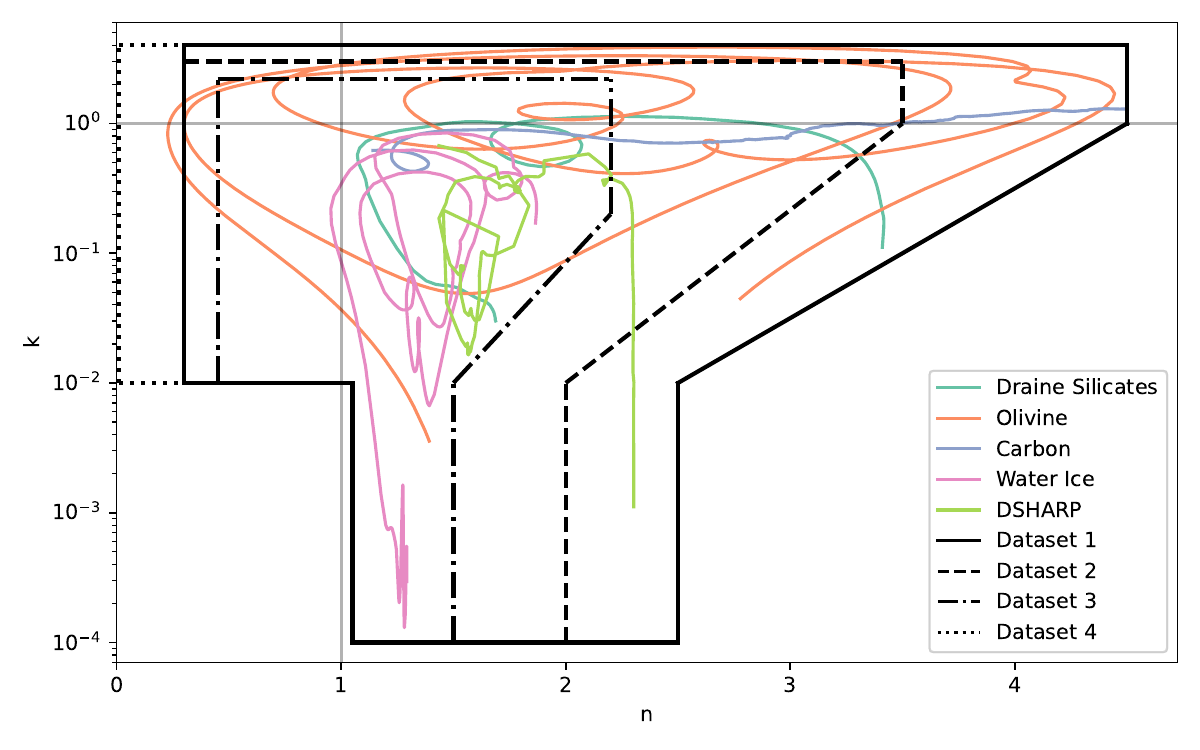}
    \caption{
        Refractive indices for a few reference compositions in comparison to the coverage in $n$ (horizontal axis) and $k$ (vertical axis) for different datasets (see Table~\ref{tab:dataset_description} for details).
        Each composition covers a different wavelength range. 
        ``Draine Silicates" refers to the astrosilicates from \cite{Draine2003ARA&A..41..241D} covering from 0.6~$\mu$m to 400~$\mu$m. ``Olivine" refers to crystalline San Carlos olivine at 300~K along the $z$-axis covering 6.7~$\mu$m to 50~$\mu$m \citep{Zeidler2015ApJ...798..125Z}. 
        ``Carbon" is the amorphous carbon from \cite{Zubko1996MNRAS.282.1321Z} from 0.1~$\mu$m to 50~$\mu$m.
        ``Water Ice" corresponds to water ice from \cite{Warren2008JGRD..11314220W}. 
        ``DSHARP" is the composition from \cite{Birnstiel2018ApJ...869L..45B} shown from 0.1~$\mu$m to 9~mm. 
       The vertical and horizontal gray lines mark $n=1$ and $k=1$, respectively. 
    %Draine astrosilicate: (0.6, 400), 
    %Olivine: (6.7, 50) at 300~K along the $a$-axis
    %Carbon: (0.1, 50), 
    %water ice: (1.45, 100.0)
    %DSHARP: (0.1, 7mm) 9?
    }
    \label{fig:nk_bounds}
\end{figure*}

Fig.~\ref{fig:nk_bounds} shows the distribution of $n$ and $k$ for different compositions and the boundaries in $n$ and $k$. 
The boundaries are limited based on available computational resources, since increasing $\xenc$ and $n$ increases the calculation time. 
The calculation time is generally faster when $k \sim 1$, meaning it increases with increasing $k$ when $k>1$ but also increases with decreasing $k$ when $k<1$. 
We produce different datasets depending on the $\xenc$ detailed in Table~\ref{tab:dataset_description}. 
Datasets 1a and 1b have the widest coverage in ($n,k$) with the smallest $\xenc$. 
Datasets 2a and 2b have an intermediate coverage in $(n,k)$ with $20 \leq \xenc \leq 35$. 
The lower limit of $\xenc$ is chosen to overlap with Datasets 1a and 1b. 
Dataset 3 covers $\xenc$ from 34 to 65 but has the most restricted $(n,k)$. 
Dataset 4 is a small supplemental set to cover the small region in $n<0.3$ and $k\in [10^{-2}, 4]$. This is specifically for a few crystalline species that extend beyond the coverage of Dataset 1. 
The maximum uncertainty of each scattering quantitiy for each dataset is also listed in Table~\ref{tab:dataset_description}. 

\begin{table*}
    \centering
    \begin{tabular}{c c|c c c c c c}
        \hline
        (1) & Dataset & 1a & 1b & 2a & 2b & 3 & 4 \\
        \hline
        (2) & \# & 1774 & 1154 & 339 & 340 & 300 & 100 \\
        (3) & $\xenc$ & (0.1, 25) & (0.1, 25) & (20, 35) & (20, 35) & (34, 65) & (0.1, 35) \\
        (4) & $n$ & (0.3, 4.5) & (0.3, 4.5) & (0.3, 3.5) & (0.3, 3.5) & (0.45, 2.2) & (0.01, 0.3) \\
        (5) & $k$ & ($10^{-4}$, 4) & ($10^{-4}$, 4) & ($10^{-4}$, 3) & ($10^{-4}$, 3) & ($10^{-4}$, 2.2) & (0.01, 4.0) \\
        (6) & $\Delta \theta_{\text{max}}$ [deg] & 3 & 1 & 3 & 1 & 1 & 1 \\
        (7) & $\frac{\delta \Cext}{\Cext}$ [\%] & (0.2, 1.4) & (0.2, 1.4) & (0.27, 0.71) & (0.26, 0.71) & (0.28, 0.61) & (0.23, 1.3) \\
        (8) & $\frac{\delta \Cabs}{\Cabs}$ [\%] & (0.2, 1.7) & (0.2, 1.6) & (0.26, 0.75) & (0.26, 0.75) & (0.27, 0.70) & (0.25, 1.3) \\
        (9) & $\delta \epsilon$ [\%] & 1.6 & 1.4 & 0.89 & 0.91 & 0.90 & 0.88 \\
        (10) & $\delta \omega$ [\%] & 1.0 & 1.1 & 0.31 & 0.39 & 0.35 & 1.4 \\
        (11) & $\delta N_{11}$ [\%] & 5.7 & 5.2 & 6.9 & 6.2 & 10 & 4.6 \\
        (12) & $\delta N_{12}$ [\%] & 2.0 & 1.8 & 0.98 & 0.93 & 0.95 & 0.78 \\
        (13) & $\delta N_{22}$ [\%] & 1.6 & 1.6 & 1.7 & 1.5 & 1.8 & 1.3 \\
        (14) & $\delta N_{33}$ [\%] & 1.7 & 1.8 & 1.7 & 1.6 & 1.8 & 1.3 \\
        (15) & $\delta N_{34}$ [\%] & 1.9 & 1.8 & 0.93 & 1.0 & 1.0 & 0.87 \\
        (16) & $\delta N_{44}$ [\%] & 3.3 & 3.1 & 3.3 & 3.3 & 3.6 & 2.7 \\
        \hline
    \end{tabular}
    \caption{
        Overview of the various datasets. 
        Row 1: The name of the dataset. 
        Row 2: Number of combinations of $(x,n,k)$. 
        Row 3: The overall range of the size parameter. 
        Rows 4, 5: The overall range of the real part and imaginary part of the refractive index, respectively. 
        Row 6: The maximum angular resolution in degrees. 
        Rows 7, 8: The range in fractional uncertainty of $\Cext$ and $\Cabs$, respectively.
        Rows 9, 10: The maximum uncertainty of $\epsilon$ and $\omega$, respectively. 
        Rows 11 - 16: The maximum uncertainty of $N_{11}$, $N_{12}$, $N_{22}$, $N_{33}$, $N_{34}$, and $N_{44}$, respectively. The minimum $\delta N_{22}$, $\delta N_{33}$, and $\delta N_{44}$ is $\sim 10^{-6} \%$. 
    }
    \label{tab:dataset_description}
\end{table*}

Lastly, we produce data at specific $n+ik$ for a series of $\xenc$ that are only meant for visualization in Sec.~\ref{sec:discussion} to demonstrate the $\xenc$ dependence. The specific $m$ are $1.7+0.01i$ which is representative of silicates and $2.5+i$ which is representative of more absorptive material. 
These were not used for training the neural network and serve as another consistency check on how well the neural network performs.

\section{Neural Network} \label{sec:neural_network}
The results from Sec.~\ref{sec:database} can be used as a lookup table through linear interpolation, but neural networks can learn nonlinear behavior.  
This section explains our neural-network training process and demonstrates that a trained model can outperform interpolation in accuracy and data size (we refer to e.g., \citealt{BailerJones2002adaa.conf...51B}, \citealt{Raschka2015}, \citealt{Pattanayak2017}, \citealt{Ting2025arXiv250612230T}, for a background on neural networks and its terminology, and see also, e.g., \citealt{Telkamp2022ApJ...939...73T}, \citealt{Kaeufer2023A&A...672A..30K}, \citealt{Palud2023A&A...678A.198P}, \citealt{Behrens2024ApJ...977...38B}, for other examples in astronomy). 

\subsection{Architecture}

We employ a feedforward neural network implemented using the Python package \texttt{PyTorch} to predict scattering properties from input parameters. 
Our architecture consists of an input layer that receives parameter values, a number of hidden layers ($n_h$), and an output layer.
Each hidden layer contains a number of neurons which has three components: a linear transformation, a Gaussian Error Linear Unit (GELU) activation function \citep{Hendrycks2016arXiv160608415H}, and a Monte Carlo dropout layer with dropout rate $\eta_d$ \citep{Srivastava2014}. 
For simplicity, we let the number of neurons for each hidden layer to be the same with $n_{\text{neu}}$.

Rather than using a single multi-output network, we train eight independent networks, each predicting one target quantity. 
The training targets for the $i$th sample are defined as:
\begin{align} \label{eq:y_transformation}
    y_{i,1} &\equiv \ln Q_{\text{ext}, i}^{\text{enc}} \text{, }
    &y_{i,2} &\equiv \epsilon_{i} \text{, } \nonumber \\
    y_{i,3} &\equiv \ln N_{11,i} \text{, }
    &y_{i,4} &\equiv N_{12,i} \text{, }
    &y_{i,5} &\equiv N_{22,i} \text{, } \nonumber \\
    y_{i,6} &\equiv N_{33,i} \text{, } 
    &y_{i,7} &\equiv N_{34,i}  \text{, }
    &y_{i,8} &\equiv N_{44,i} \text{. }
\end{align}
The uncertainties of the targets, $\delta y_{i}$, are transformed accordingly. 
The first two targets ($y_1$ and $y_2$) are angle-independent quantities that depend only on the three input parameters, which we consider $\ln \xenc$, $n$, and $\ln k$. 
The remaining six targets represent angle-dependent scattering-matrix elements that additionally require the scattering angle $\theta$ as input.

Each network is trained to minimize a weighted mean-squared error loss function:
\begin{align} \label{eq:loss_function}
    L_{j} = \frac{1}{N} \sum_{i=1}^{N} w_{i} ( y_{i,j} - \hat{y}_{i,j} )^{2}
\end{align}
where $\hat{y}_{i,j}$ is the model prediction of the $j$th target for the $i$th sample, $y_{i,j}$ is the true value, $N$ is the number of samples in the training set, and $w_{i,j}$ is the weight. 
The weights are proportional to the inverse square of the uncertainty, $\propto 1 / \delta y^{2}$. 
However, $\delta y$ can be extremely low (for example, at $\theta=0^{\circ}$ where $y_{4}$ should be 0 and $\delta y_{4} \rightarrow 0$), making the training process susceptible to numerical noise and the loss function to be dominated by a few high-weight samples. 
To mitigate this issue, we implement weights as:
\begin{align} \label{eq:weights_from_dy}
    w_{i,j} = \bigg[ \frac{ 1 }{ \text{max}( \delta y_{i,j}, 10^{-4}) } \bigg]^{2},
\end{align}
which means we treat regions with $\delta y < 10^{-4}$ with equal weights as those with $\delta y$ at the threshold. This prevents numerical instabilities and maintains appropriate relative weighting for higher-uncertainty regions.

We impose physical constraints to the model. 
While the logarithmic transformation for $y_{1}$ and $y_{3}$ guarantees that the predicted values for $\Cext$ and $Z_{11}$ are positive, we impose a sigmoid constraint to the output layer for $y_{2}$ to limit it within $[0,1]$. For the remaining targets ($y_4$ through $y_8$), we apply a $\tanh$ function limiting all predictions to the interval $[-1,1]$, which is consistent with the expected physical bounds for these normalized scattering-matrix elements.

\subsection{Training, Cross-validation, \& Hyperparameter Optimization}
% learning_rate, hidden_dims, n_neurons, epochs, dropout rate

% namely the learning rate $\zeta_{l}$, the number of neurons for each layer $n_{\text{neu}}$, the number of hidden layers $n_{\text{hid}}$, the Monte Carlo dropout rate $\zeta_{mc}$, the number of training epochs $n_{\text{epoch}}$, and the batch size. 

Training is performed using the \texttt{PyTorch}'s \texttt{AdamW} optimizer with default momentum parameters ($\beta_{1}=0.9$, $\beta_{2}=0.999$), and we fix the weight decay to $0.01$. 
We leave the learning rate $\zeta_{lr}$ as a free hyperparameter. 
The input parameters of the training data is scaled through \texttt{MinMaxScaling} from \texttt{scikit-learn}.
The number of training epochs is also left as a free hyperparameter. 

%There are a number of hyperparameters from our adopted architecture. 
Since optimal hyperparameter combinations are typically problem specific and difficult to determine a priori, we conduct a systematic grid search for each of the eight target networks. 
We withhold 10\% of the data as test data to evaluate the final optimized hyperparameter. The remaining 90\% is used as training and validation through $k$-fold ($k=5$) cross-validation using \texttt{KFold} from \texttt{scikit-learn} to ensure robust model evaluation. 
Noise in the validation loss over epochs complicates comparison across hyperparameters. 
Since the validation loss can exhibit fluctuations over epochs, we smooth over each curve with a uniform kernel with size 11, then average across the five folds as the representative validation curve. 
We take the standard deviation across the five folds to assess model stability. 

For each hyperparameter, we train a maximum number of epochs of 2020 for $y_{1}$ and $y_{2}$. Since training $y_{3}$ to $y_{8}$ takes more time, we train only to maximum epochs of 1020 and tested fewer combinations of $n_{\text{neu}}$ and $n_{\text{hid}}$. Table~\ref{tab:hyperparameter_search} lists the grid of hyperparameter values we tested. From $y_{1}$ and $y_{2}$, we found that including Monte Carlo dropout does not improve the model and thus we leave $\zeta_{d}=0$ for training the angular quantities. In order to determine the optimal number of epochs, we compare the loss at intervals of 50 epochs starting from 100 as opposed to taking the direct minimum along epochs, which is susceptible to fluctuations.

When choosing the optimal hyperparameter set, we consider all the sets with scores that are within 1 standard deviation of the lowest score and adopt the set with the smallest standard deviation. 
Table~\ref{tab:best_hypar} shows the optimal hyperparameter for each target.

\begin{table}
    \centering
    \begin{tabular}{l l}
        \hline 
        $y_{1}$ and $y_{2}$ \\
        \hline  
        $n_{\text{hid}}$ & 2, 4, 6, 8, 10 \\
        $n_{\text{neu}}$ & 20, 50, 100, 150 \\
        $\zeta_{lr}$ & $\expnumber{1}{-5}$, $\expnumber{3.2}{-5}$, $\expnumber{1}{-4}$, $\expnumber{3.2}{-4}$, $\expnumber{1}{-3}$, $\expnumber{3.2}{-3}$ \\
        Batch Size & 32, 64, 128 \\
        $\zeta_{d}$ & 0, 0.2 \\
        Epochs & 100 to 2000 in steps of 50 \\
        \hline 
        $y_{3}$ to $y_{8}$ \\
        \hline 
        $n_{\text{hid}}$ & 4, 6, 8 \\
        $n_{\text{neu}}$ & 100, 200, 300 \\
        $\zeta_{lr}$ & $\expnumber{1}{-5}$, $\expnumber{3.2}{-5}$, $\expnumber{1}{-4}$, $\expnumber{3.2}{-4}$, $\expnumber{1}{-3}$, $\expnumber{3.2}{-3}$ \\
        Batch Size & 128, 256 \\
        $\zeta_{d}$ & 0 \\
        Epochs & 100 to 1000 in steps of 50 \\
        \hline 
    \end{tabular}
    \caption{
        The hyperparameters explored for each target. 
    }
    \label{tab:hyperparameter_search}
\end{table}

\begin{table}
    \centering
    \begin{tabular}{l l l l l l}
        \hline
        Target & Batch Size & $\zeta_{lr} $ & $n_{\text{hid}}$ & $n_{\text{neu}}$ & Epochs \\
        \hline 
        $y_{1}$ & 64 & $\expnumber{3.2}{-4}$ & 4 & 150 & 1500 \\
        $y_{2}$ & 64 & $\expnumber{3.2}{-4}$ & 4 & 150 & 1650 \\
        $y_{3}$ & 256 & $\expnumber{1}{-4}$ & 6 & 300 & 700 \\
        $y_{4}$ & 256 & $\expnumber{3.2}{-5}$ & 8 & 300 & 800 \\
        $y_{5}$ & 256 & $\expnumber{1}{-4}$ & 8 & 300 & 700 \\
        $y_{6}$ & 128 & $\expnumber{3.2}{-5}$ & 6 & 300 & 950 \\
        $y_{7}$ & 256 & $\expnumber{1}{-4}$ & 8 & 300 & 950 \\
        $y_{8}$ & 128 & $\expnumber{3.2}{-5}$ & 6 & 300 & 750 \\
        \hline
    \end{tabular}
    \caption{
        The optimal hyperparameter set for each target. 
        We found that the optimal $\zeta_{d}$ is $0$ for both $y_{1}$ and $y_{2}$. 
    }
    \label{tab:best_hypar}
\end{table}

\subsection{Results}

%For $\Qextvol$ and $\epsilon$, we find that the best-fit hyperparameter has $n_{\text{free}} \sim 5$ times more than the number of data. This resonates with the commonly known phenomena where overly parameterized .... \red{cite}. 

\begin{figure*}
    \centering
    \includegraphics[width=\textwidth]{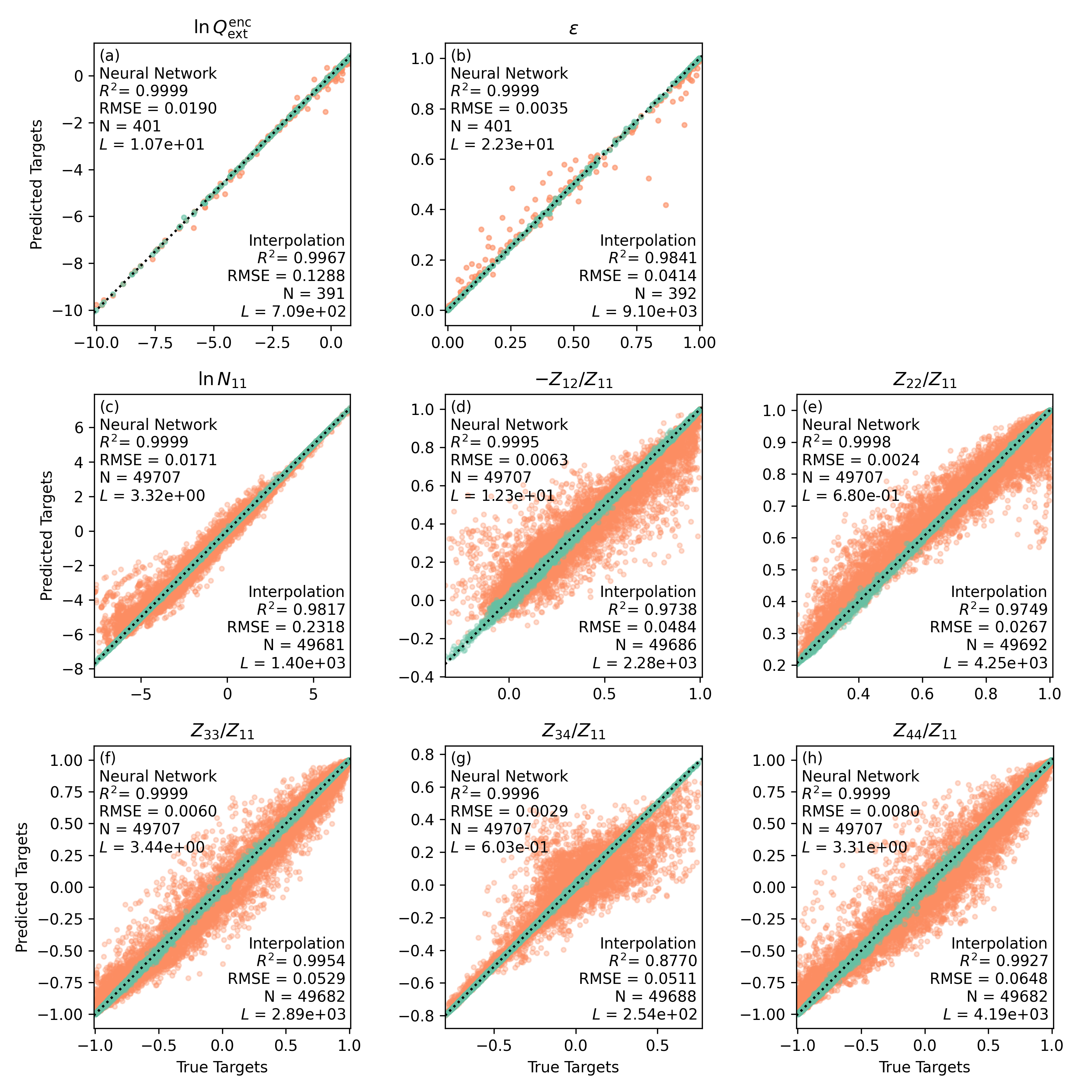}
    \caption{
        Predictions of the neural network (green) and linear interpolation (orange) against the test data.
        The neural network clearly outperforms the linear interpolation in accuracy. 
        The horizontal axes show the true value and the vertical axes are the corresponding predicted values. 
        The dashed black lines show the ideal case of perfect prediction. 
        The text in each panel shows the performance of the neural network and linear interpolation, where $R^{2}$, $L$, RMSE, and $N$ refer to the coefficient of determination, the value of the loss function (Eq.~\ref{eq:loss_function}), the root mean squared error, and the number of points used for the assessment. 
    }
    \label{fig:pred_vs_true}
\end{figure*}

Plotting the model predictions versus the true values is a useful way to visualize accuracy, shown in Fig.~\ref{fig:pred_vs_true} for each of the 8 targets. 
A perfect prediction should lie along the diagonal. 
We also list common metrics like the Coefficient of Determination ($R^{2}$) and the Root Mean Squared Error (RMSE) defined by
\begin{align}
    \text{RMSE} = \sqrt{ \frac{1}{N} \sum_{i=1}^{N} ( y_{i,j} - \hat{y}_{i,j} )^{2} }
\end{align}
for each $j$th target. RMSE is particularly beneficial since it has the same units as the training targets. 
%Relevant performance metrics like $L$ (Eq.~\ref{eq:loss_function}), $R^{2}$, and RMSE, are shown along with Fig.~\ref{fig:pred_vs_true}. 
The relevant metrics are listed in Table~\ref{tab:performance}. 
%Table~\ref{tab:performance} lists additional metrics, like the $\bar{\Delta}$ and ${\Delta}_{\text{max}}$. 

We compare the results with linear interpolation using the same test data. 
We use the \texttt{LinearNDInterpolator} from \texttt{scipy} to conduct interpolation of the unstructured dataset. 
For equal comparison, we interpolate in $y$, which applies the target transformations using Eq.~\ref{eq:y_transformation}. 
Linear interpolation cannot occur if the sampling point is beyond the convex hull of the input data. Thus, we ignore those data points when assessing the accuracy. 
The performance is also listed in Fig.~\ref{fig:pred_vs_true} and Table~\ref{tab:performance}. 
%Note that when measuring the speed, the linear interpolation requires an initial setup time that is not part of the measured time. While the setup time is small for the angle-independent quantities, that time is not negligeable for the scattering-matrix elements and takes around a couple hours with our training data. 

Clearly, the neural networks offer much lower $L$ values (Fig.~\ref{fig:pred_vs_true}), which makes sense, since linear interpolation does not take into account the data uncertainty. 
Even if we compare RMSE, which does not take into account data uncertainty, neural network models also outperforms interpolation. 
Overall, we find that the trained neural network is more accurate than linear interpolation across all quantities, as expected given its well-known ability to capture non-linear behavior. 

%We compare the speed by averaging the execution time over the number of samples. The speed for the angular quantities are consistently faster than interpolation by $10^{2}$ to $10^{3}$ times (Table~\ref{tab:performance}), while being slower than interpolation for the angle-independent quantities by a factor of 3 to 4. We note that the time for linear interpolation can vary depending on how linear interpolation is implemented (e.g., on a grid). The time comparison here is meaningful specifically for linear interpolation on unstructured data. 

For a sense of calculation speed, we time \texttt{glitterin} on a Macbook pro (M3 chip). For all eight quantities, it takes $\sim 0.7$~milliseconds for a single $\theta$ point and $\sim 5$~milliseconds for 181 $\theta$ points at a single $\xenc$, $n$, and $k$. The final size of the models is $13$~MB.

\begin{table}
    \centering
    \begin{tabular}{r c c c}
        \hline 
        Target & Model & $L$ & RMSE \\
        (1) & (2) & (3) & (4) \\
        \hline
        $y_{1}$ & NN & 10.7 & 0.0190 \\
            & Interp. & $7.09 \times 10^{2}$ & 0.1288 \\
        \hline 
        $y_{2}$ & NN & 22.3 & 0.0035 \\
            & Interp. & $9.10 \times 10^{3}$ & 0.0414 \\
        \hline
        $y_{3}$ & NN & 3.32 & 0.0171 \\
            & Interp. & $1.40 \times 10^{3}$ & 0.2318 \\
        \hline 
        $y_{4}$ & NN & 12.3 & 0.0063  \\
            & Interp. & $2.28 \times 10^{3}$ & 0.0484 \\
        \hline 
        $y_{5}$ & NN & 0.68 & 0.0024  \\
            & Interp. & $4.25 \times 10^{3}$ & 0.0267  \\
        \hline
        $y_{6}$ & NN & 3.44 & 0.0060 \\
            & Interp. & $2.89 \times 10^{3}$ & 0.0529  \\
        \hline 
        $y_{7}$ & NN & 0.603 & 0.0029 \\
            & Interp. & $2.54 \times 10^{2}$ & 0.0511 \\
        \hline 
        $y_{8}$ & NN & 3.31 & 0.0080  \\
            & Interp. & $4.19 \times 10^{3}$ & 0.0648  \\
        \hline
    \end{tabular}
    \caption{
        Some performance metrics of the neural network and its linear interpolation counterpart. 
        See Fig.~\ref{fig:pred_vs_true} for the other performance metrics. 
        Column 1: The target. 
        Column 2: The neural-network model or the linear interpolation case. 
        Column 3: The value of the loss function (Eq.~\ref{eq:loss_function}). 
        Column 4: The root mean squared error. 
        %Column 5: The average difference $\Delta$  defined in the appropriate physical space (Eq.~\ref{eq:Delta_j}) of the test data for each target. 
        %Column 6: The maximum $\Delta$ of the test data for each target.
    }
    \label{tab:performance}
\end{table}

\section{Discussion} \label{sec:discussion}

In this section, we demonstrate the irregular grain calculations as estimated by the neural network, compare results against spherical grains, and discuss the impacts on dust inferences. 

\subsection{Cross-Sections} \label{sec:cross_sections}

\begin{figure*}
    \centering
    \includegraphics[width=\textwidth, trim={0.4cm 0.5cm 0.2cm 1.2cm}, clip]{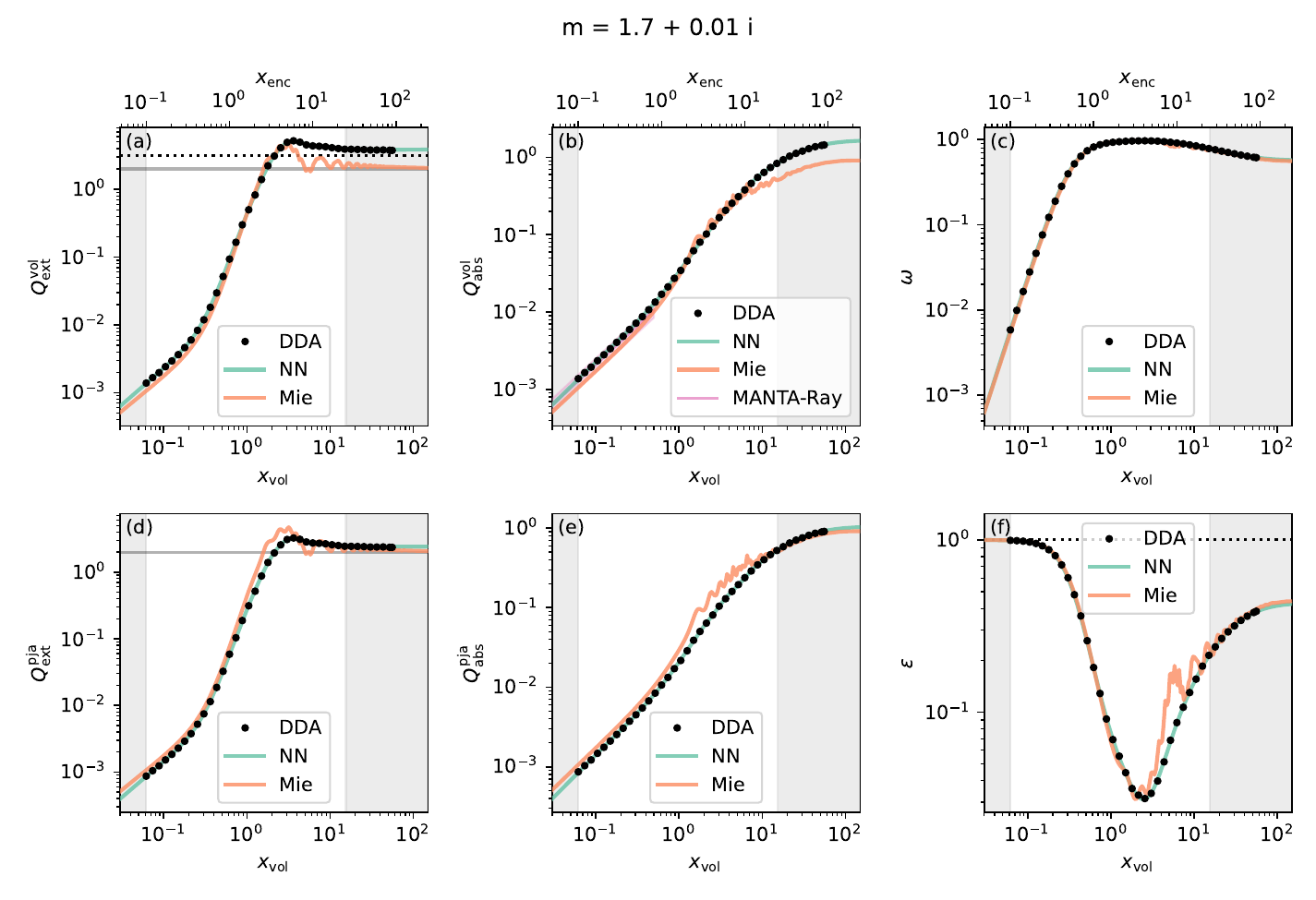}
    \caption{
        Extinction and absorption efficiencies and the albedo as a function of $\xvol$ for irregularly shaped grains and spherical grains of the same volume and refractive index $m=1.7+0.01i$. 
        Panels a and d: The exinction efficiencies $\Qextvol$ and $\Qextpja$ (see Eq.~\ref{eq:C_eq_Q_pi_r2}). 
        The values for spheres are the same in both panels. 
        %The offset results from fundamental differences in the morphologies of the grains. 
        Panels b and e: The absorption efficiencies $\Qabsvol$ and $\Qabspja$. 
        We additionally show estimates from \texttt{MANTA-Ray} in panel b where $\Qabsvol$ is available. 
        The shaded region shows $20\%$ error. 
        Panels c and f: The albedo $\omega$ and $\epsilon$. 
        For each panel, the green solid lines are predictions from the neural network. 
        The dotted points are independent DDA calculations that the neural network never trained on. The datapoint of the largest grain we could simulate has $\xenc=90$ ($\xvol \sim 56$). 
        The orange lines are calculations from Lorenz-Mie theory. 
        %The dark blue curve corresponds to the dipole approximation. 
        %The dipole overlaps with Lorenz-Mie calculations as $\xvol \rightarrow 0$. 
        The top axes show $\xenc$ for the irregularly shaped grains. 
        The left and right regions are shaded to show the minimum and maximum $\xenc$ of the training data for the neural network at $m=1.7+0.01i$. 
        We show predictions of the neural network beyond this range to visualize how well it generalizes to size parameters beyond its training data. 
    }
    \label{fig:plot_x_1}
\end{figure*}

Fig.~\ref{fig:plot_x_1} plots $\Qextvol$, $\Qabsvol$, $\omega$, and $\epsilon$ as a function of $\xvol$ with $n=1.7$ and $k=0.01$ for the neural network and Lorenz-Mie theory. 
%As a comparison, we also include the spherical dipole case, which should approximate the small-size-parameter regime \citep{vandeHulst1957lssp.book.....V, Bohren1983asls.book.....B}. 
We further include results of irregularly shaped grains at various $\xenc$, which were calculated separately from the datasets (the neural networks were never trained on these independent calculations). 
We also include a comparison with \texttt{MANTA-Ray}, which provides a fitted analytical function to $\Qabsvol$ for small, irregularly shaped grains produced by \cite{Lodge2024MNRAS.535.1964L}. The fractal dimension ($d_{f}$) is a unique input as part of the grain geometry considered in \texttt{MANTA-Ray}, and we adopt $d_{f}=2.1$ as a representative value. To compare with our results, we use an uncertainty of $20\%$ based on their reported fitting accuracy and structural differences. 

Fig.~\ref{fig:plot_x_1}a shows the rise of $\Qextvol$ at small size parameters, $x < 1$, which is linearly proportional to $x$ for irregularly shaped grains and spherical grains: the extinction cross-section is proportional to the volume of the particle. The two cases are comparable, since the volumes are equivalent and the refractive index is relatively small. However, at $\xvol \sim 5$, $\Qextvol$ reaches a maximum and asymptotically approaches a constant as $\Cext$ approaches the geometric cross-section in the geometric-optics regime. 
From the extinction paradox, we always expect $\Qextpja=2$ for any grain. 
Indeed, Fig.~\ref{fig:plot_x_1}d shows $\Qextpja$ for both shapes and we see that the irregularly shaped grains approach $2$ at larger $\xvol$. 
For spherical grains, $\Qextvol=\Qextpja$ simply because $\rvol=\rpja$ and approach $\Qextvol=2$ in the geometrical limit. However, for an irregularly shaped grain, $\Qextvol$ approaches a constant at a value greater than $2$. 
This difference results because irregularly shaped grain casts a larger projected area than the spherical grain with the same amount of material ($\rvol < \rpja$). 
The discrepancy highlights a fundamental difference between irregularly shaped grains and spherical grains: the extended morphology alone significantly impacts the cross-sectional quantities. 
One can derive that 
\begin{subequations}
\begin{align}
    \Qextvol &\rightarrow 2 g f^{-2/3} \\
    \Qextenc &\rightarrow 2 g
\end{align}
\end{subequations}
when $\Qextpja \rightarrow 2$ from the extinction paradox. 
In other words, the limiting behavior of $\Qextvol$ is completely determined by the structural parameters $f$ and $g$.
With our adopted $f$ and $g$, we would expect $\Qextvol \rightarrow 3.200$ in the geometric regime. 
%Given that the $\Qextpja$ profile of the irregular grain has yet to reach $2$, our DDA calculations here have not approached the geometric-optics limit, but they have captured the transition region. 

It is convenient to first look at the albedo $\omega$ in Fig.~\ref{fig:plot_x_1}c, f. 
Intriguingly, $\omega$ (and also $\epsilon$) appears similar between spheres and irregularly shaped grains in this case. 
Both $\epsilon$ from irregularly shaped grains and from spheres shows a dip near $\xvol \sim 2$ and then an increase to approximately the same value towards larger $\xvol$, although the spherical case exhibits several oscillations in the intermediate regime. Evidently, the fractional amount of light that is absorbed versus scattered does not depend sensitively on particle morphology with the adopted refractive index. 

Since $\epsilon$ is mostly similar between both shapes, the difference of $\Qabsvol$ shown in Fig.~\ref{fig:plot_x_1}b mostly tracks the difference of $\Qextvol$. The irregularly shaped grains have similar $\Qabsvol$ as the spherical grains for small $\xvol$. Indeed, the result is consistent with \texttt{MANTA-Ray}. 
Deviations occur for $\xvol$ beyond $\sim 5$ as $\Qextvol$ begins to track the projected geometric area. 
The values in this regime show a clear enhancement of $\Qabsvol$ compared to the values from spheres that are due to the more extended geometry of the irregularly shaped grains than spheres for the same amount of material. 
%As we will see, higher $m$ produces enhanced absorption even at small size parameters. 

The neural-network predictions also shown in Fig.~\ref{fig:plot_x_1} demonstrates successful approximations to the independent DDA calculations. 
Note that the neural network only outputs $\Qextenc$ and $\epsilon$, while $\Qabsenc$ is derived. 
An intriguing result from the neural network is that it can extrapolate at $\xenc < 0.1$, which is beyond the boundaries of the training set. Extrapolation in this regime, however, is not particularly significant, since small size parameters are easy for DDA to calculate. 
%Errors begin to increase as the size goes below $\xenc =0.1$. The uncertainty of $\omega$ in the training data is approximately $\sim 1\%$ percent, which means values below that are more susceptible to averaging error. 
% extrapolation 
The more interesting case is extrapolation to larger $\xvol$. 
The independent calculation goes to $\xenc=90$ ($\xvol \sim 56$) and we find that the fractional difference with the neural-network prediction of $\Qextenc$ is $2\%$, which is comparable to the uncertainty due to ensemble averaging. 
The absolute difference for $\epsilon$ is 0.002 at $\xenc=90$, which is also appreciably small. 
The extrapolation for $\Qextenc$ is not very useful to some degree, since we know the geometric-optics limit, but the accuracy for $\epsilon$ is quite surprising and potentially useful. 
We note that $\Qextenc$ at values beyond $\xenc \sim 150$ does not appear to converge towards the expected limit which is a clear indicator of limitations in extrapolation much beyond parameter space of the training data. Thus, values there should be treated with care.

\begin{figure*}
    \centering
    \includegraphics[width=\textwidth, trim={0.4cm 0.5cm 0.2cm 1.2cm}, clip]{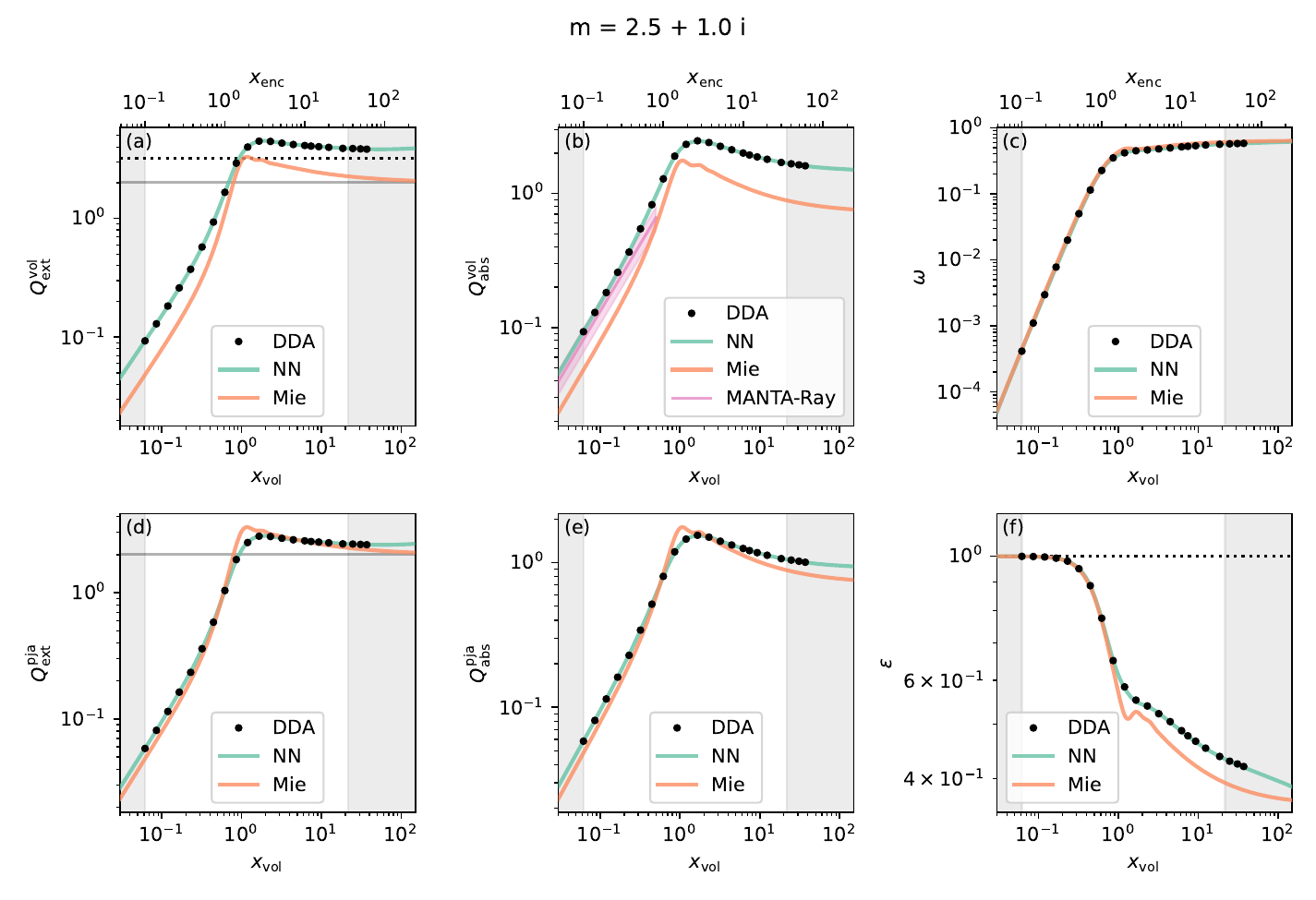}
    \caption{
        The extinction efficiency and absorption efficiencies, and the albedo as a function of $\xvol$ for $m=2.5+1i$. 
        The figure is plotted in the same way as Fig.~\ref{fig:plot_x_2}. 
        Panel b shows that irregularly shaped grains have a larger absorption cross-section than a sphere having the same material volume in the small grain regime ($\xvol<1$) and large grain regime ($\xvol>10$). 
    }
    \label{fig:plot_x_2}
\end{figure*}

We present another example at $m=2.5 + i$ as a demonstration of a higher $m$ case, shown in Fig.~\ref{fig:plot_x_2} where we independently calculate the irregularly shaped grains at multiple $\xenc$ for comparison, up to $\xenc=60$. 
Fig.~\ref{fig:plot_x_2}a and d show the $\Qextvol$ and $\Qextpja$. Towards the geometric limit ($\xvol \gg 1$), we once again see that $\Qextpja \rightarrow 2$ for both spheres and irregularly shaped grains, which leads to $\Qextvol > 2$ for the irregularly shaped grains, as in the previous case with $m=1.7+0.01i$. 
A new difference is in the small $\xvol$ regime, where $\Qextvol$ of irregularly shaped grains are not equal to that of spheres. 
As seen in Fig.~\ref{fig:plot_x_2}b, the difference is due to higher $m$ for small $\xvol$, which has been reported in previous studies \citep[e.g.][]{Lodge2024MNRAS.535.1964L}. Indeed, the results shown here are consistent with the estimation from \texttt{MANTA-Ray}. 
Fig.~\ref{fig:plot_x_2}c and f show the $\omega$ and $\epsilon$. We see that $\omega$ remains similar as in the previous example, but only at small $\xvol$ with the higher $m$. A bigger difference in the $\epsilon$ between spheres and irregularly shaped grains appears when $\xvol > 1$.

%The geometrical effect should impact both $\Qextvol$ and $\Qabsvol$ equally meaning it does not impact $\epsilon$. An increased $\epsilon$ of irregularly shaped grains demonstrates that an enhanced absorption effect also exists for large $\xvol$. Whether this originates from the effect as that for small $\xvol$ is unclear. 

\subsection{The 10~$\mu$m Complex} \label{sec:10_micron_complex}

\begin{figure*}
    \centering
    \includegraphics[width=\textwidth]{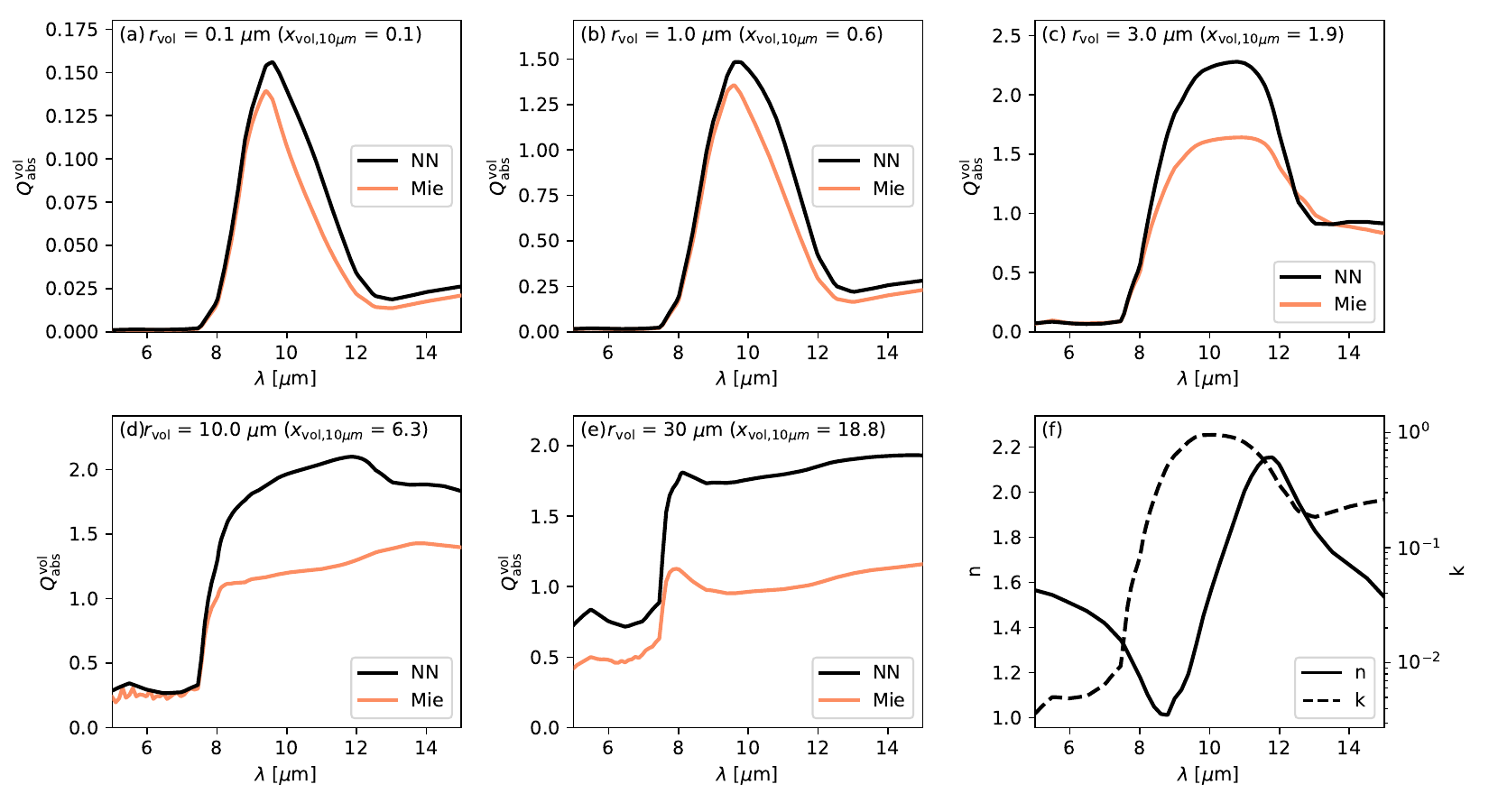}
    \caption{
        Panel a-e: Absorption efficiency of amorphous pyroxene grains near the 10~$\mu$m region with $\rvol=0.1$, 1, 3, 10, and 30~$\mu$m. 
        The results for spherical grains are shown in green, and the neural-network results are in black. 
        Regions where the neural network is extrapolation beyond the $\xenc$ of its training data is plotted as dotted lines (only in panel e). 
        $\Qabsvol$ gives the cross-section for grains of the same material volume (Eq.~\ref{eq:C_eq_Q_pi_r2}).
        Panel f: The $n$ (solid lines) and $k$ (dashed line) values.
    }
    \label{fig:nvm_asize_pyroxene}
\end{figure*}

Silicate features are often used to infer the dust-formation conditions and composition in disks (Sec.~\ref{sec:intro}).
In this subsection, we compare $\Qabsvol$ in the 10~$\mu$m silicate complex produced by spheres with that produced by irregularly shaped grains. 
We first explore amorphous pyroxene for different grain sizes in Fig.~\ref{fig:nvm_asize_pyroxene} where Fig.~\ref{fig:nvm_asize_pyroxene}f shows the adopted $n+ik$ across $\lambda$ \citep{Dorschner1995A&A...300..503D}.
Fig.~\ref{fig:nvm_asize_pyroxene}a shows the small grain case, where $\xvol < 1$ across wavelength, and we see that $\Qabsvol$ is similar except between 10 to 12~$\mu$m where $\Qabsvol$ of irregularly shaped grains is enhanced due to a higher $m$. 

As the grain sizes increase, the contrast between the feature peak and its wings begins to diminish for both sets of grains, which is a simple result of approaching the geometrical limit for both morphologies. 
However, the shape of the profiles and how it changes to reach that limit differ for the two cases. 
For the 30~$\mu$m grains in Fig.~\ref{fig:nvm_asize_pyroxene}e, $\Qabsvol$ is $\sim 1.1$ at $\lambda > 10$~$\mu$m for spheres, but $\sim 1.6$ for irregularly shaped grains. 
As mentioned in Sec.~\ref{sec:cross_sections}, the irregularly shaped grains provide a larger cross-section than spheres of the same material volume as the grains move beyond $\xvol \sim 1$. At $\lambda < 10$~$\mu$m (where $\xvol > 18.8$), $\Qabsvol$ of irregularly shaped grains is greater than that of spheres, while at $\lambda > 10$~$\mu$m ($\xvol < 18.8$), $\Qabsvol$ is even more enhanced in part due to the geometrical effect and the high $k$ as discussed above.

\begin{figure*}
    \centering
    \includegraphics[width=\linewidth]{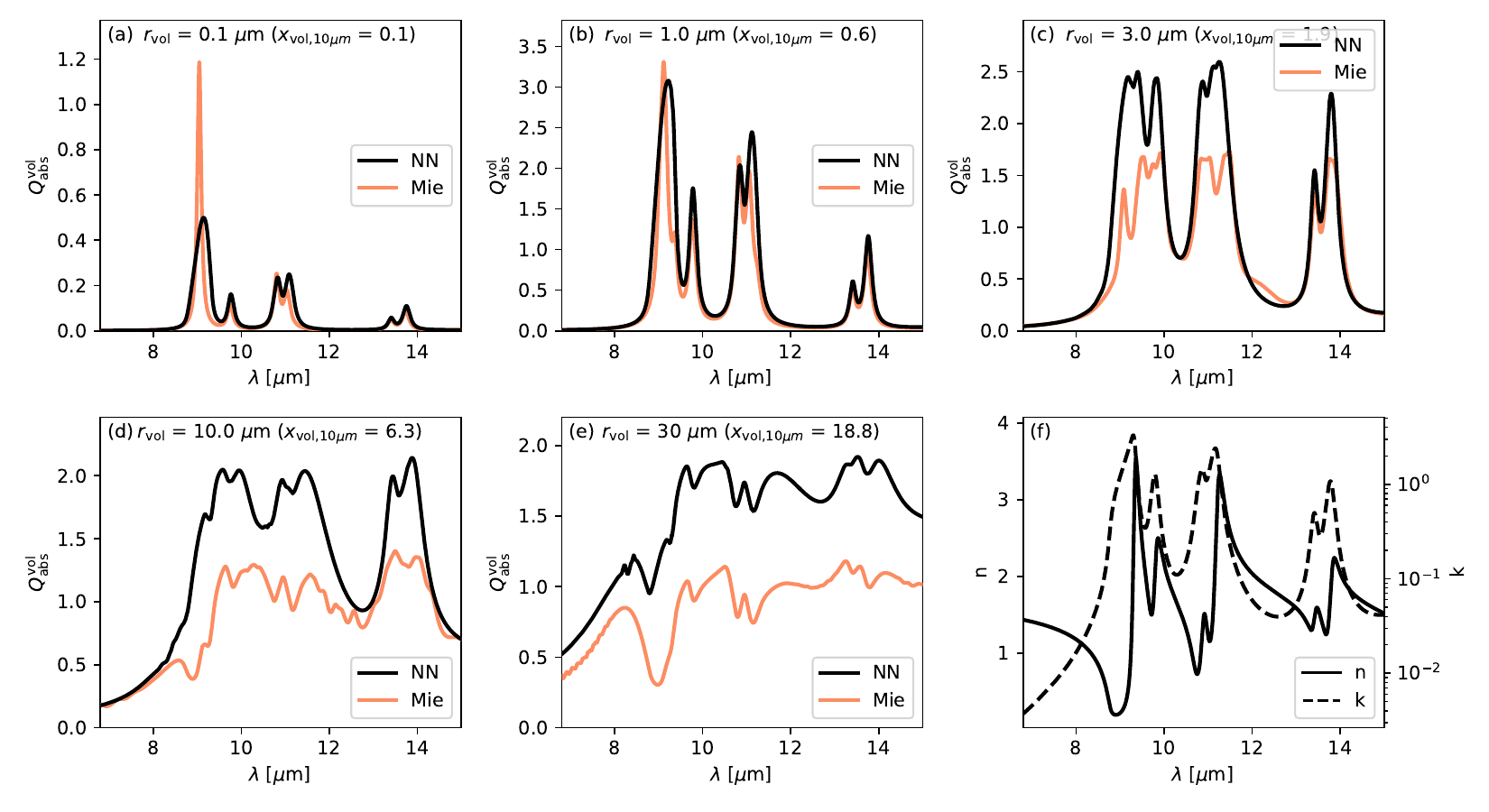}
    \caption{
        The absorption efficiency $\Qabsvol$ for crystalline enstatite plotted in the same way as Fig.~\ref{fig:nvm_asize_pyroxene}. 
    }
    \label{fig:nvm_asize_enstatite}
\end{figure*}

As another example, we look at crystalline enstatite to study how the finer solid-state features of a crystalline species are impacted by grain shape. 
The refractive index is shown in Fig.~\ref{fig:nvm_asize_enstatite}f \citep{Jaeger1994A&A...292..641J} and Fig.~\ref{fig:nvm_asize_enstatite}a through e show grains of various $\rvol$. 
For $\rvol = 0.1$~$\mu$m (Fig.~\ref{fig:nvm_asize_enstatite}a), the spectra are mostly similar, but the features at $\lambda=9$ and 11~$\mu$m for irregularly shaped grains are both enhanced relative to the spherical case also due to a higher $m$.
Nevertheless, for $\rvol=0.1$, 1, and 3~$\mu$m (Fig.~\ref{fig:nvm_asize_enstatite}a-c), one can identify similar locations of the fine feature. 
However, for $\rvol=10$ and 30~$\mu$m grains (Fig.~\ref{fig:nvm_asize_enstatite}d, e), the spectrum of spherical grains bear little resemblence to that of the irregularly shaped grains. 
Beyond the specific peak locations, the biggest difference is in the systematic enhancement of irregularly shaped grains at $\lambda > 9$~$\mu$m. The broad offset is once again due to the more geometrically extended nature of irregular grains, while the enhancement due to high $m$ leads to more complicated wavelength dependence.

The systematic enhancement of $\Qabsvol$ of irregularly shaped grains compared to spherical grains could easily impact dust inferences. 
Modeling assuming spherical grains would underpredict emission of the large grains, since the spherical grains are too compact compared to their irregular counterparts having the same amount of material. This could be the reason why modeling debris disks sometimes requires a blackbody component or a removal of the dust continuum when modeling the mid-IR solid state features \citep[e.g.][]{Weinberger2011ApJ...726...72W, Lu2022ApJ...933...54L}. 
A differential underprediction of emission of large grains would also impact size distribution inferences. 
Furthermore, the differences in the detailed locations of the peaks of the solid-state features could affect inferences of gas emission \citep[e.g.][]{Kanwar2024A&A...689A.231K}. 

\subsection{Scattering-Matrix Elements} 
% debris disk: optical/IR, total intensity and polarization 
% good recovery
% evidence of extrapolation 

\begin{figure*}
    \centering
    \includegraphics[width=\textwidth,trim={0.3cm 0.5cm 0 1.2cm},clip]{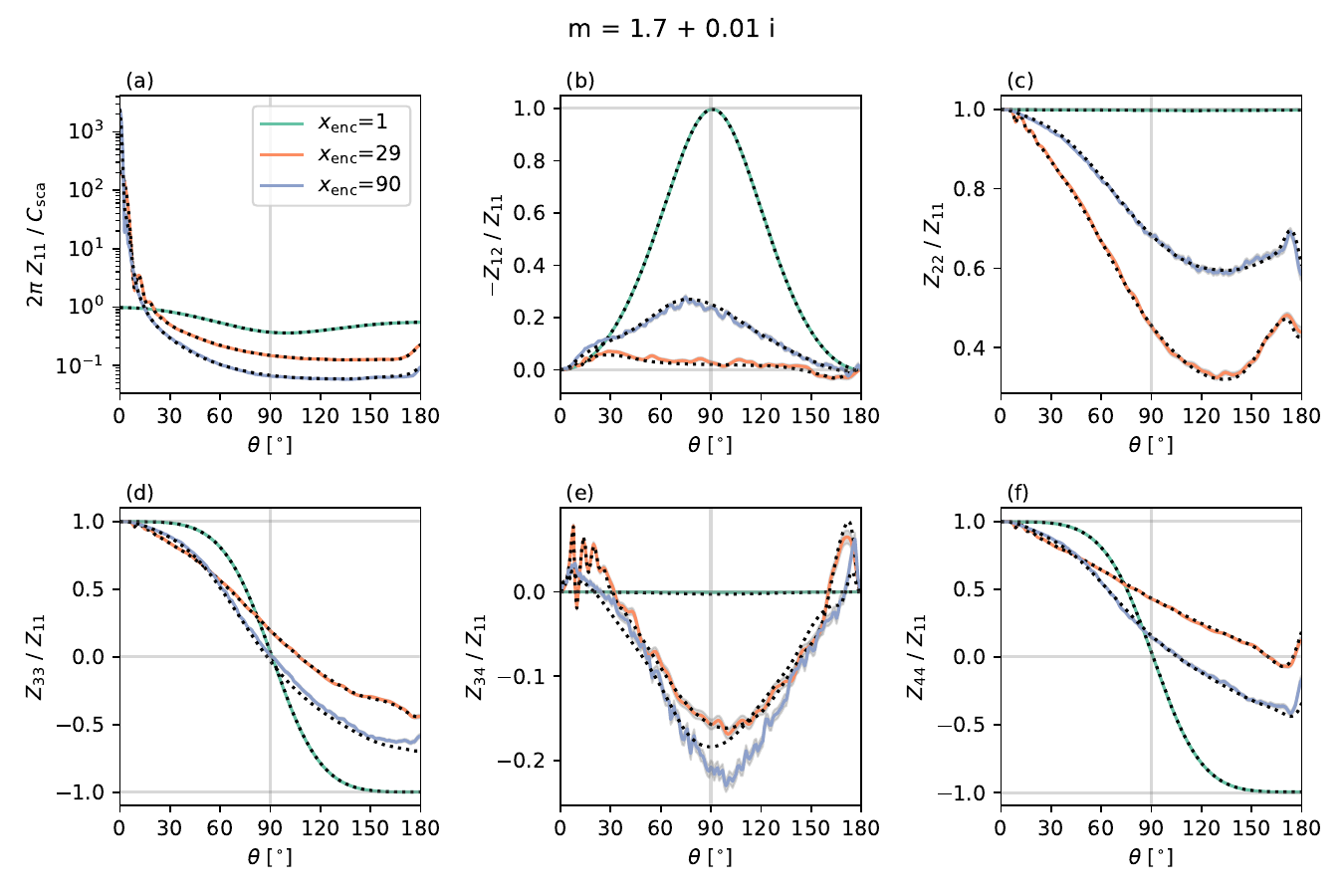}
    \caption{
        Comparing the neural network against an independently calculated visualization set with $m=1.7+0.01i$. 
        Each panel shows one of the scattering-matrix elements for $\xenc=1, 29$, and $90$, in green, orange, and blue solid lines, respectively. 
        The filled grey region corresponds to the $1\sigma$ error in the population average. 
        The dotted black lines denote the neural-network results for each $\xenc$ and are not colored for easier interpretation, since the model values are all close to the independent calculations. 
        The horizontal grey lines in panels b, c, d, e, and f denote $N_{ij}=-1$, $0$, or $1$ and the vertical grey lines denote $\theta=90^{\circ}$. 
    }
    \label{fig:zij_1.7_0.01i}
\end{figure*}

The scattering-matrix elements of dust and its connection to the grain size and composition is key to interpreting the several resolved total intensity and polarization images of debris disks in the optical (see Sec.~\ref{sec:intro}). We first compare the neural network against the independent, visualization sets.  
Fig.~\ref{fig:zij_1.7_0.01i} compares the scattering profiles using the $m=1.7+0.01i$ visualization data at different $\xenc$. We show $\xenc=1$ and $29$, which are within the parameter space in the training data. 
Clearly, the neural network successfully approximates the matrix elements and the level of differences are comparable to the uncertainty of the data. 

The model works surprisingly well even at our largest independent DDA simulation at $\xenc=90$. As a reference, the RMSE for $y_{3}$ to $y_{8}$ are $0.12$, $0.012$, $0.0084$, $0.037$, $0.024$, and $0.030$, respectively (with $181$ points in $\theta$).  
At this $m$, the largest $\xenc$ in the training data is $35$ and in fact, $\xenc=90$ is larger than the $\xenc$ in the entire dataset (across all $m$), which means the neural network is extrapolating to size parameters beyond what it was trained on. 
The independent DDA simulation at $\xenc=90$ took on average $\sim 113$~hours per sample on our machine, which means 6.4~years for 500 samples in total to obtain the ensemble-averaged scattering quantities (we only completed 410~samples here). 
The trained neural network only takes $\sim 5$~milliseconds for the ensemble averaged value (for all 8 targets using the same number of $\theta$ points), representing a $10^{10.6}$ times speed up. 

\begin{figure*}
    \centering
    \includegraphics[width=\textwidth,trim={0.3cm 0.5cm 0 1.2cm},clip]{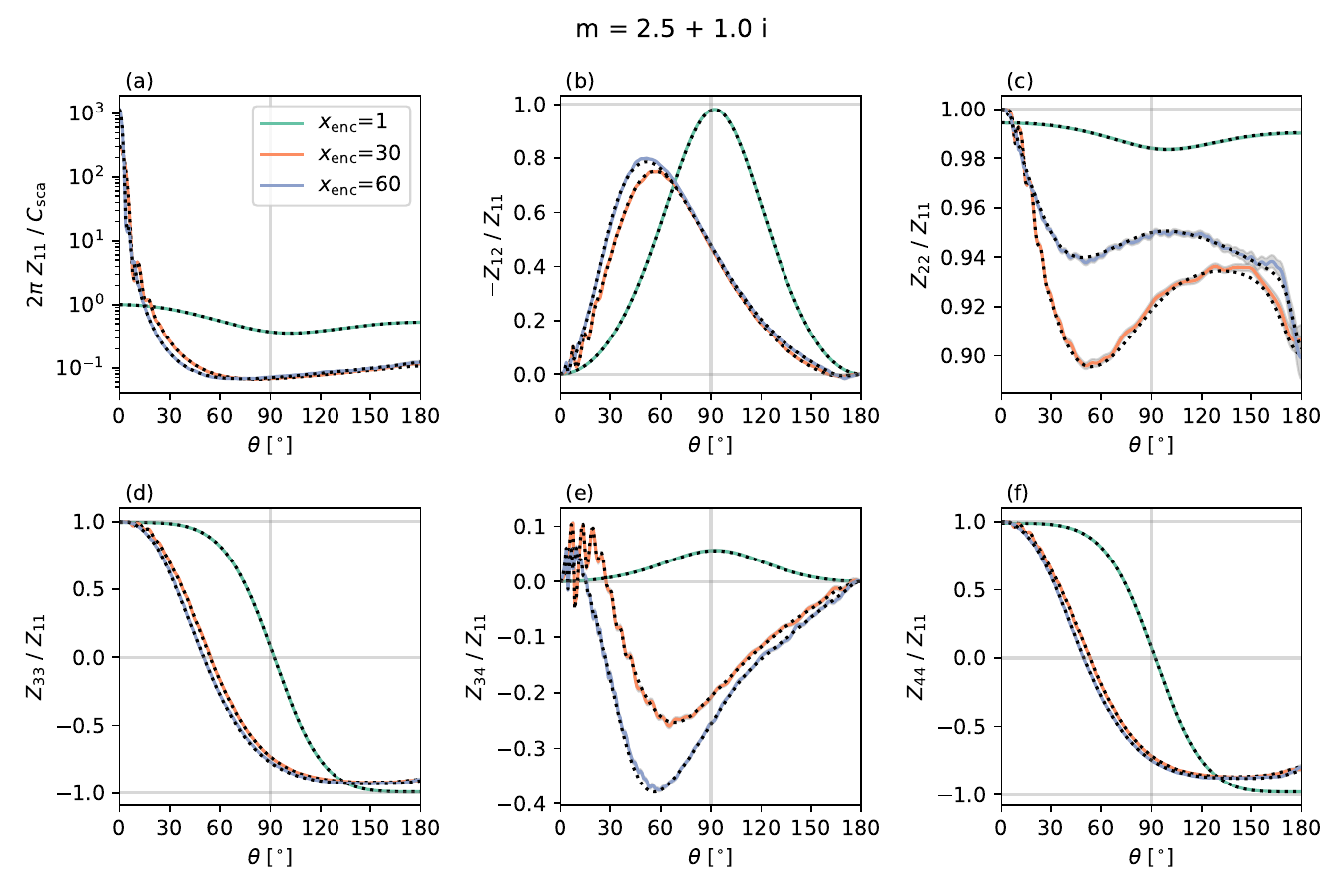}
    \caption{
        The scattering-matrix elements for $m=2.5+i$ plotted in the same way as Fig.~\ref{fig:zij_1.7_0.01i}.
    }
    \label{fig:zij_2.5_1.00i}
\end{figure*}

As another example, Fig.~\ref{fig:zij_2.5_1.00i} shows the scattering profiles using the higher $m$ case with $m=2.5+i$.
The maximum $\xenc$ of the training set at this $m$ is $35$. Once again, we observe that the model is able to approximate the profiles that are within its parameter space ($\xenc=1$, $30$) and also beyond its parameter space ($\xenc=60$). 
The RMSE of $y_{3}$ to $y_{8}$ at $\xenc=60$ are $0.030$, $0.0070$, $0.0015$, $0.0060$, $0.0056$, and $0.0055$, respectively (with 181 points in $\theta$). 
The average calculation time per sample for $\xenc=60$ is $\sim 11$ hours on our machine ($8$~months for 500 samples), or a $10^{9.6}$ times speed up by $\texttt{glitterin}$. 

However, we caution that it is difficult to assess the accuracy of extrapolation. 
If we could produce simulations at those large size parameters, we would utilize those into the training and aim to extrapolate to even broader parameter spaces. 
Nevertheless, this demonstrates that while interpolation has a very strict boundary in the input parameter space, the trained neural network has a softened boundary. 
We suspect that the neural network was able to learn the trends of scattering that do not vary quickly with changing parameters. 
%For example, the scattering-matrix elements between $\xenc=30$ and $\xenc=90$ differ, but are not drastically beyond the realm of possibilities. This is consistent with the well known capability of neural networks (\red{cite}). 
The potential to reduce computation time by ten billion times at the cost of $\sim 10\%$ accuracy in extrapolation is very attractive and should enable more rapid development and exploration in the data analysis. 
Still, any use of extrapolation beyond the limits of the training data should be treated with caution.

%The results shown here should help model optical/NIR images of debris disks where the single scattering of stellar light directly impacts how the resolved total intensity images appear (\red{cite}). 
%The $Z_{12}$ element will help interpret the polarized intensity images (\red{cite}) in a uniform manner across sources and simultaneously offer insight in the dust properties . 

% trim={<left> <lower> <right> <upper>}

\subsection{Comparison to Laboratory Data} \label{sec:scattering_lab_data}
% good representation of real grains

With a well-trained model in hand, the biggest question is whether the model can approximate actual grains.  
We explore the laboratory-measured scattering-matrix elements from the Granada-Amsterdam Light Scattering Database for suitable data \citep{Munoz2025JQSRT.33109252M}. 
We first compare the results for feldspar \citep{Volten2001JGR...10617375V}. 
A similar comparison was done by \cite{Zubko2013JQSRT.131..175Z} using agglomerated debris particles. 
Thus, a comparison here also serves as a consistency check. 

The laboratory measurements include estimates of the size distribution using a laser particle sizer \citep{Munoz2021ApJS..256...17M}.
How one converts between the size parameter in this paper to the size parameter inferred from the laboratory measurements is ambiguous due to different methods in defining particle sizes. Presumably, it should differ only by a constant factor of order unity and the relative size distribution should be more reliable. 
In addition, the size-distribution data has a lower limit, since the smallest sizes are difficult to capture experimentally. 
Detailed fitting of the laboratory data is beyond the scope of this paper and we demonstrate the capability of \texttt{glitterin} as is.

We also compare predictions by Lorenz-Mie theory.
The measured size distribution from the laboratory is too discrete for Lorenz-Mie theory, which produces significant oscillations as a function of scattering angle. Thus, we increase the number of points by a factor of 10 in the size distribution through linear interpolation to produce smoother curves in the scattering matrix. We use the same upsampled size distribution for the neural network for equal comparison, but we find that the scattering matrix does not differ much before and after upsampling for the irregularly shaped grains. 

Since the laboratory $Z_{11}$ is a normalized quantity, we also normalize the $Z_{11}$ curves for the neural network and Lorenz-Mie results. 
We normalize $Z_{11}$ to $\theta=90^{\circ}$, instead of the default $\theta=30^{\circ}$ from the laboratory, since it is more convenient for observations of debris disks, which better probe side-scattering \citep[e.g.][]{Arriaga2020AJ....160...79A, Kueny2024ApJ...961...77K}. 
Also, our $Z_{34}$ differs from the laboratory by a negative sign due to a difference in convention. 
Here we modify the laboratory $Z_{34}$ by a negative sign to maintain consistency within the paper. 

We use the measurements of feldspar at $\lambda=441.6$~nm \citep{Volten2001JGR...10617375V}. 
For the refractive index, we adopt a nominal value of $n=1.5$ and $k=10^{-4}$ given that $n$ and $k$ are estimated to be in the range $1.5 \leq n \leq 1.6$ and $10^{-5} \leq k \leq 10^{-3}$, respectively. 
The effective particle size estimated from the Fraunhofer method is $1$~$\mu$m which corresponds to a size parameter of 14.

\begin{figure*}
    \centering
    \includegraphics[width=\textwidth]{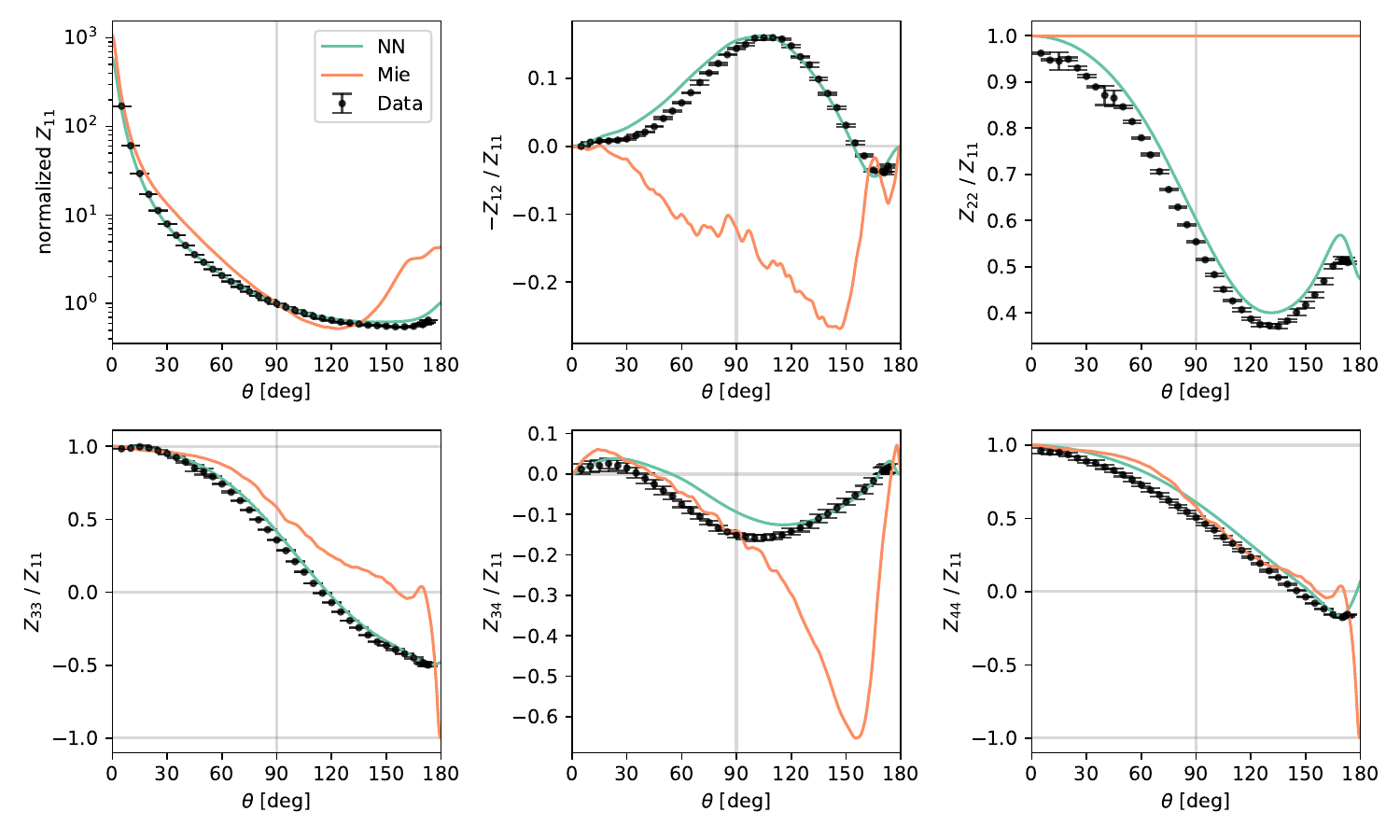}
    \caption{
        The non-zero scattering-matrix elements of forsterite particles measured from the laboratory (black dots) compared to predictions from spherical grains (solid orange line) and those from the neural network (solid green line) using the same size distribution and $m=1.5+10^{-4}i$. 
        Panel a: the $Z_{11}$ element normalized at $\theta=90^{\circ}$. 
        The neural network trained on the agglomerated debris particles better captures the scattering properties of real, irregularly shaped grains than Lorenz-Mie theory. 
    }
    \label{fig:Nij_feldspar}
\end{figure*}

Fig.~\ref{fig:Nij_feldspar} shows the comparison between the laboratory measurements, results from the neural network, and those from Lorenz-Mie theory. 
Clearly, the neural-network results are much more similar to the laboratory measurements than the spherical counterpart across all scattering-matrix elements. 
Spherical grains show a systematic discrepancy compared to irregularly shaped grains. For example, the prominent backscattering of $Z_{11}$ (Fig.~\ref{fig:Nij_feldspar}a), the difference in sign of $N_{12}$ (Fig.~\ref{fig:Nij_feldspar}b), a constant $N_{22}=1$ (Fig.~\ref{fig:Nij_feldspar}c), strong backscattering of $N_{34}$ Fig.~\ref{fig:Nij_feldspar}e), and $N_{33}=N_{44} \rightarrow -1$ as $\theta \rightarrow 180^{\circ}$ (Fig.~\ref{fig:Nij_feldspar}d, f). 
Our results are consistent with those of \cite{Zubko1996MNRAS.282.1321Z}. 

As a second demonstration, we use hematite (Fe$_{2}$O$_{3}$) measured at $\lambda=632.8$~nm \citep{Shkuratov2004JQSRT..88..267S, Munoz2006A&A...446..525M}. 
The mineral is of particular interest because of its high $n$ and $k$, which is the highest available in the light-scattering database, and allows us to test the performance of the neural network in this regime. 
Hematite is a birefringent material with $n=2.9$ and $n=3.1$ for the extraordinary and ordinary axes, respectively, while the imaginary part is in the range of $10^{-2} \leq k \leq 10^{-1}$ \citep{Sokolik1999JGR...104.9423S}. We adopt $n=3.0$ and $k=0.03$ as nominal values for our calculations. 
Note that the refractive index is beyond the boundaries in the training data. 
The effective particle size estimated from the Fraunhofer method is $0.4$~$\mu$m, meaning the size parameter is $\sim 4$.

\begin{figure*}
    \centering
    \includegraphics[width=\textwidth]{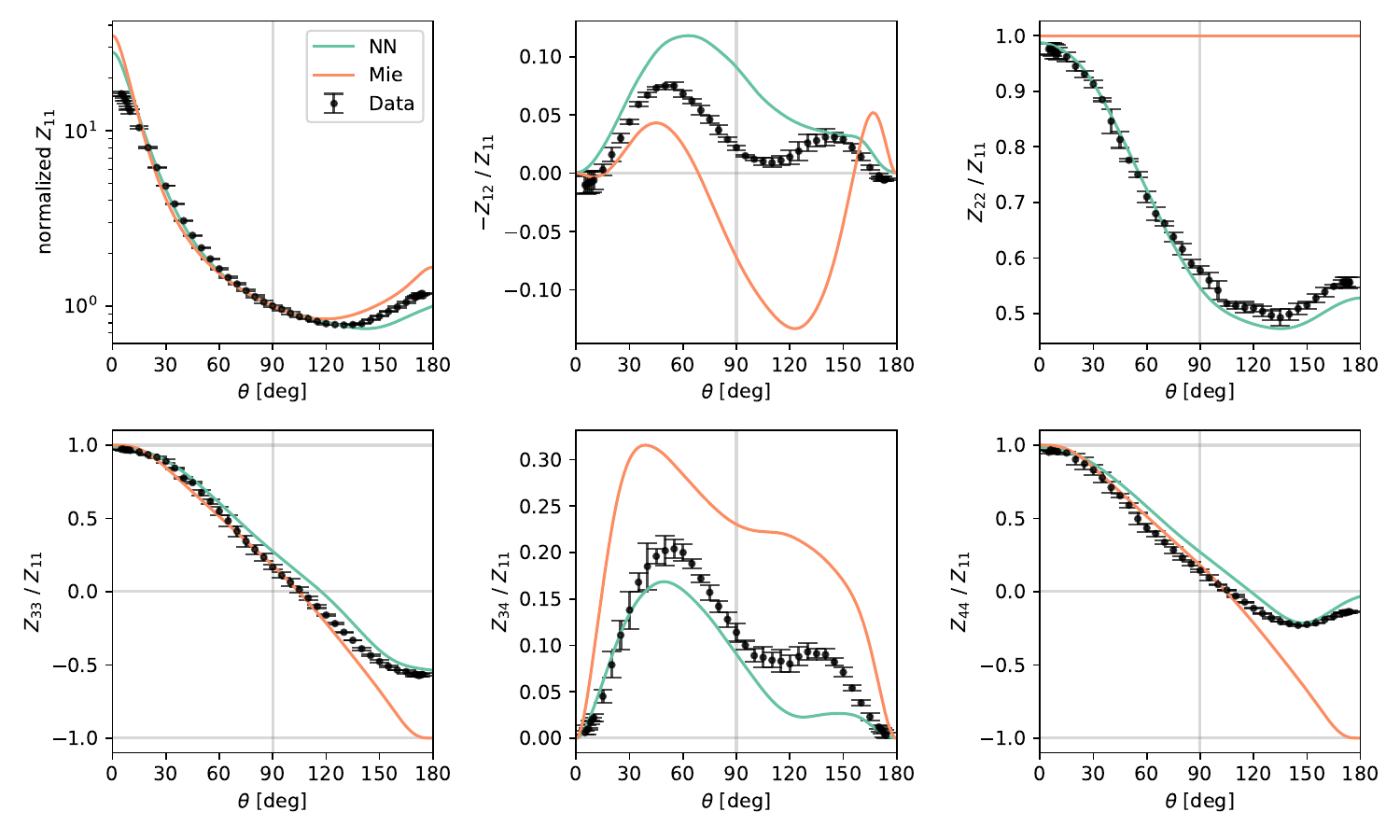}
    \caption{
        Similar to Fig.~\ref{fig:Nij_hematite}, but for hematite particles at $\lambda=632.8$~nm.
        The adopted refractive index is $m=3+0.03i$. 
    }
    \label{fig:Nij_hematite}
\end{figure*}

Fig.~\ref{fig:Nij_hematite} shows the comparisons and we find once again that the neural network is a better approximation of the laboratory measurements than the Lorenz-Mie theory. 
Elements $N_{12}$ and $N_{34}$ show discrepancies, but the neural network is at least able to capture the qualitative behavior that the spherical grains cannot, similar to the feldspar case above. 
Most of the differences between the laboratory measurements and the neural network are within $\sim 5\%$, which is surprising given the uncertainty in the size distribution, the artificial accomodation for birefringence, the extrapolation beyond the boundaries of $m$, and the differences in the underlying shape of the grains. 

The two examples here demonstrate that the neural network trained on agglomerated debris particles offers a much better approximation to the scattering of real, irregularly shaped grains, achieving successful qualitative behavior and similar quantitiative behavior over two distinctly different regimes of the refractive index. 
The differences are at around $\sim 5\%$. 
More detailed comparisons may require careful extrapolation in the size distribution \citep{Zubko2013JQSRT.131..175Z} or considering the uncertainty of the refractive index, which is beyond the scope of this paper. 
Nevertheless, a qualitatively accurate and fast model with even a conservative precision of $\sim 10\%$ is very attractive. 
In contrast, Lorenz-Mie theory is a precise model for spherical grains, but not accurate in characterizing realistic grains. 

The hematite demonstration here marks the fourth such example where structures like agglomerated debris particles have successfully approximated laboratory measured scattering-matrix elements with the previous samples being feldspar \citep{Zubko2013JQSRT.131..175Z}, forsterite \citep{Zubko2024JQSRT.32309053Z}, and olivine \citep{Videen2018JQSRT.211..123V}. 

\subsection{Emission and Self-Scattering Polarization in the Millimeter Wavelength Regime} \label{sec:millimeter_wavelength}

Another light-scattering application from irregularly shaped grains is in interpreting the ALMA images at millimeter wavelengths. 
ALMA has been routinely providing resolved observations of debris disks \citep[e.g.][]{MacGregor2017ApJ...842....8M}. 
%However, it has been difficult to model optical images and millimeter wavelengths simultaneously using a single population of spherical grains \citep{Rodigas2015ApJ...798...96R}. 
For protoplanetary disks, where the optical depth is much higher, multiple scattering of thermal photons can affect the observed flux and polarization \citep{Miotello2023ASPC..534..501M}. 
When modeling with spherical grains, the dust sizes inferred from the millimeter-wave emission are too large to accurately replicate the measured polarization, making a single solution difficult to obtain \citep[e.g.][]{Lin2020MNRAS.496..169L, Zhang2023ApJ...953...96Z}.  
Both environments require proper characterization of the emission and scattering of realistic, irregularly shaped grains. 

Here we define the cross-section per unit mass, or simply the opacity, in units of cm$^{2}$ per gram as
\begin{align}
    \kappa_{i} \equiv \frac{ C_{i} }{ \frac{4\pi}{3} \rho_{s} \rvol^{3} }
\end{align}
where $i$ can be ``ext," ``abs," or ``sca" for the extinction, absorption, and scattering, respectively. The denominator corresponds to the mass of the grain, where $\rho_{s}$ is the specific weight of the grain material.
We use the default DSHARP composition as a demonstration and adopt their $\rho_{s}=1.675$~g cm$^{-3}$. 
For reference, at $\lambda=0.87$, $1$, $3.1$, and $7$~mm, $n \sim 2.3$ and $k\sim 2.3 \times 10^{-2}, 2.1 \times 10^{-2}, 8.4 \times 10^{-3}$, and $4.3 \times 10^{-3}$, respectively. 

For direct comparison with literature, we consider a size distribution where the number of particles with radius within $r$ to $r+dr$ is $n(r)dr$ and follows $n(r) \propto r^{-3.5}$. 
The minimum grain size (by $\rvol$) is fixed at $0.1$~$\mu$m and we vary the maximum grain size. 
Since the $0.1$~$\mu$m grains at millimeter wavelengths have a much smaller $\xvol$ than what the neural network was trained on, we assess the opacity at $\xenc=0.1$ and extrapolate to smaller sizes by scaling through $\Cabs \propto \xenc^{3}$ and $\Csca \propto \xenc^{6}$. 
The frequency dependence of dust $\kabs$ is often used to constrain the grain size \citep[e.g.][]{Hildebrand1983QJRAS..24..267H, Draine2006ApJ...636.1114D} and is defined as $\beta \equiv \partial \kabs / \partial \nu $.

\begin{figure*}
    \centering
    \includegraphics[width=\textwidth]{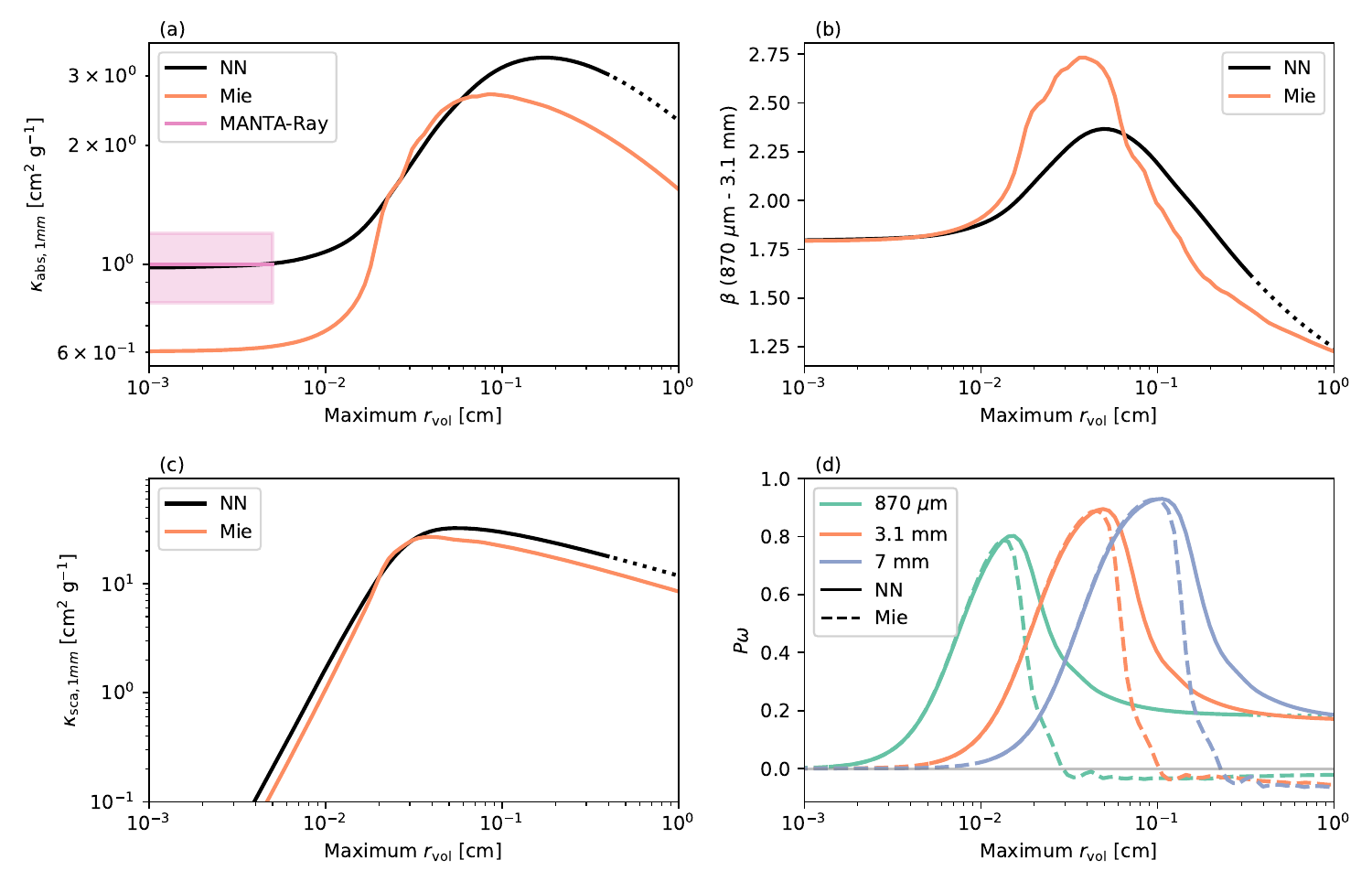}
    \caption{
        Differences in the emission and scattering behavior between spheres and irregularly shaped grains in the millimeter wavelength regime. 
        Panel a and b: the absorption and scattering opacity ($\kabs$ and $\ksca$) at $\lambda=1$~mm using the DSHARP composition for a size distribution of grains with different maximum $\rvol$. 
        The black lines are the neural-network results where the dotted section are results where the maximum $\rvol$ exceeds the maximum size parameter in the training set (extrapolation). 
        The green line shows the results obtained using Lorenz-Mie theory. 
        In panel a, the blue line is the \texttt{MANTA-Ray} result in the small grain limit and the shaded region represents a 20\% uncertainty. 
        Panel c: the opacity index $\beta$ between $\lambda=$870~$\mu$m and 1.3~mm. 
        Panel d: the albedo multiplied by the degree of linear polarization at $\theta = 90^{\circ}$. $P\omega$ is an assessment of the level of polarization from self-scattering. 
        Different colors correspond to different wavelengths. The dashed lines are results for spheres. The solid lines are results from the neural network where its dotted sections mark the extrapolation regions. 
        The horizontal grey line marks $P\omega=0$. 
    }
    \label{fig:kappa_amax}
\end{figure*}

Fig.~\ref{fig:kappa_amax}a and c show $\kabs$ and $\ksca$ at $\lambda=1$~mm. For $\kabs$, we also include the result from \texttt{MANTA-Ray} and demonstrate again that the $\kabs$ of irregularly shaped grains in our study is similar to that of $\texttt{MANTA-Ray}$ when $\xvol < 1$ \citep{Lodge2024MNRAS.535.1964L}. 
At maximum $\rvol > 1$~mm, both $\kabs$ and $\ksca$ of the irregularly shaped grains are greater than those of the spherical grains, which is a result of a larger cross-section per amount of material from grains with $\xvol > 1$ (see Sec.~\ref{sec:cross_sections}).  
Note that beyond maximum $\rvol > 2.5$~mm, the size-averaged opacity from the neural network starts to include extrapolation to large size parameters and should be treated with caution.
An enhanced absorption opacity in the millimeter regime would impact estimations of the dust mass using the millimeter flux density \citep[e.g.][]{Miotello2023ASPC..534..501M}.
However, overall, $\kabs$ and $\ksca$ between irregularly shaped and spherical grains do not vary by more than an order of magnitude with the adopted refractive index. 

As a result of the change in the shape of $\kabs$, $\beta$ is different from the spherical case with a more diminished peak and broader extent (Fig.~\ref{fig:kappa_amax}b).
A similar change was shown for porous grains approximated by the effective medium theory \citep{Tazaki2018ApJ...860...79T, Tazaki2019ApJ...885...52T}. 
The difference could potentially impact inferences of the size distribution \citep[e.g.][]{Macias2021A&A...648A..33M, Sierra2021ApJS..257...14S, Zhang2023ApJ...953...96Z, Doi2023ApJ...957...11D}.

The greatest impact of morphology is in the polarization. 
We assess polarization by defining $P \equiv - Z_{12}(\theta=90^{\circ}) / Z_{11}(\theta=90^{\circ})$ and compare the product, $P \omega$, at different $\lambda$.
Fig.~\ref{fig:kappa_amax}d compares $P\omega$ from the neural network and spherical grains at various $\lambda$. 
The discrepancy is obvious when grains are larger than the observing wavelength. 
Using $\lambda=870$~$\mu$m as an example, the spherical case becomes a negative when the maximum grain size is $0.03$~cm or greater, but the polarization for the irregularly shaped grains remains entirely positive. 
The behavior is consistent with other approximation methods \citep{Tazaki2019ApJ...885...52T}. 
The negative sign means a $90^{\circ}$ difference in the polarization angle and has been used to infer maximum sizes of $\sim 100$~$\mu$m when adopting spherical grains \citep{Yang2016MNRAS.456.2794Y}. 
A fast emulator of irregularly shaped grains could be the solution to the discrepencies in grain sizes inferred from multiwavelength total intensity and polarization \citep[e.g.][]{Ohashi2020ApJ...900...81O, Harrison2024ApJ...967...40H, Zhang2023ApJ...953...96Z, Lin2024MNRAS.528..843L} and also support proper analysis of polarization surveys and address grain growth beyond single-source characterizations. 

\subsection{Caveats and Outlook}
% 
% grain structure:
% - the process can be adopted for other grain structures
% porosity and refractive index coverage
% - we can expand this model
% 

In this exploratory work, we propose that we can use neural networks trained on DDA simulations of irregular grains as a solution to the longstanding computational barrier of retrieving dust information from astronomical environments. 
With the capability to approximate realistic light-scattering properties from irregularly shaped grains in a fraction of the time as numerical methods, we believe the model makes it feasible to replace Lorenz-Mie theory with tools that achieve more realistic results. 
There are a number of caveats in this demonstration that require caution in its interpretation. 

First, we have adopted the agglomerated debris particles prescription as our grain structure. 
As demonstrated in Sec.~\ref{sec:cross_sections}, the morphology, which determines $f$ and $g$, changes the limiting behavior of $\Qextvol$ and $\Qabsvol$. 
Different morphologies, like different levels of porosity, should also change the cross-sections for the same volume of material. 
We can also expect changes to the scattering-matrix elements, since the angular dependence is determined by the hierarchical structure at different spatial scales \citep[e.g.][]{Tazaki2016ApJ...823...70T}. 
Nevertheless, empirical explorations suggest that the differences between irregular grain morphologies is likely not very large \citep{Zubko2015JQSRT.150...42Z, Zubko2024JQSRT.32309053Z}, at least not large compared to using spheres. 
With fast computational speed, we envision that it might be possible to use observations to differentiate the effects of particle morphology.

Second, our coverage in $m$ in this work is limited mostly as a balance between computational time and we focus on species of interest because higher values of $m$ become challenging for DDA. The current dataset is not wide enough to cover all of the widely adopted dust species, like troilite, but the training process here is expansive. With more computational resources, these challenges can be overcome. In the future, we plan to incorporate more samples and release newer versions of the neural network. 

Another related note is we have assumed homogenous and isotropic material. 
While some light-scattering properties of inhomogeneous grains can be approximated by mixing rules to a certain degree \citep{Videen2015JQSRT.150...68V}, the situation is less clear for anisotropic material (birefringent or trirefringent material). 
At least in our exploration of randomly oriented, birefringent hematite (Sec.~\ref{sec:scattering_lab_data}), any systematic discrepancy from adopting an isotropic material is perhaps not very large. 

%Lastly, one may ask whether it is actually beneficial in considering neural networks versus simply adding more data points to lookup tables of scattering quantities. 
%Certainly, it is preferable to remove the training error using results directly from DDA calculations if there are sufficient computational resources. 
%However, we can observe that the human time necessary for the analysis of astronomical data in addition to radiative-transfer modeling is already challenging, making it difficult to go one step further in conducting computationally intensive DDA calculations for each new project. 
%This is likely why much of the target-specific dust modeling in astrophysics has been confined to using Lorenz-Mie and effective medium theories.
%I don't think I agree or understand the following sentence as you've provided counterexamples, and also seem to counter it in the sentence immediately following: 
%Neural networks are a natural improvement to any lookup table by learning the nonlinear behavior. 
%since it can outperform a database in accuracy (or equivalently, less demand for the number of data points) and the scaled down size of the product makes downloading easier allowing more widespread use and more memory efficient for users in general. 
%For angle-independent quantities, which only depend on 3 parameters, the benefit is likely not large. Linear interpolation is fast and accurate enough with additional points, and the size of the data is manageable. The clearest benefit only emerges when training on 4 parameters. 
%For example, in our training datasize is \red{.... }

Given the computational speed and better light-scattering performance to real particles, we believe that the neural network is the next step in replacing the role of Lorenz-Mie theory in conducting dust inferences in debris disks and other astronomical environments. 
Considering the cost of DDA simulations, it is an open question of how to most efficiently produce training data for other grain structures for future neural networks. 
The training data and visualization data are publicly available to the community to explore more efficient and precise neural networks beyond what is explored here \footnote{See \url{https://doi.org/10.7910/DVN/QZE3D7} and \url{https://doi.org/10.7910/DVN/CWLYZL}.}

\section{Conclusion} \label{sec:conclusion}

Using ADDA, we simulate light scattering of irregularly shaped grains defined through agglomerated debris particles for a wide range of complex refractive index $m$ and size parameter $x$.  
We introduce \texttt{glitterin}, which is a neural network trained on the simulated data. 
The goal of this paper is to produce one possible neural network that is both accurate and fast and provides a practical pathway to larger scale analysis of observations and better radiation-transfer simulations without being plagued by the assumption of spherical morphology. 
Our main conclusions are summarized as follows.

\begin{enumerate}
    \item \texttt{glitterin} is more accurate than linear interpolation, has a smaller data volume than the original data, and operates at millisecond timescales. In addition, we discover empirical cases where the neural network can generalize beyond the parameter boundaries of the original training data. The neural network and training data are made publicly available.\footnote{The neural network is available at \url{https://doi.org/10.7910/DVN/STER2G} and can be accessed using the python interace at \url{https://github.com/zheyudaniellin/glitterin}. }
    
    \item The extinction and absorption cross-sections, $\Cext$ and $\Cabs$, of irregular grains are different compared to the spherical grains with the same mass of material. When $x \ll 1$, the $\Cext$ is comparable for relatively small $|m|$. However, as $x \gg 1$, $\Cext$ of irregular grains is greater than $\Cext$ of spheres, because irregularly shaped grains are more geometrically extended than spheres for the same mass. $\Cabs$ follows a similar behavior as $\Cext$, but it can be more enhanced even for small $x$ when $m$ is large. We find that the albedos for both morphologies are comparable. The different $x$ and $m$ dependence of $\Cabs$ leads to different profile shapes of the solid-state features in the mid-IR. 

    \item We use laboratory-measured light-scattering profiles from the Granada-Amsterdam Light Scattering Database to test how well \texttt{glitterin} can approximate light scattering of realistic, irregularly shaped grains. Adopting the laboratory-inferred size distribution and refractive index of feldspar and hematite, we find the light-scattering profiles from \texttt{glitterin} resemble the laboratory-measured light-scattering behavior with discrepancies of $\sim 5\%$ in both cases. In all cases, they are a great improvement over the assumption of spherical grains, which can have enormous errors qualitatively. 

    \item We consider the mass opacity for a size distribution of grains in the millimeter wavelength range. We find that the absorption and scattering opacity of spherical grains are comparable to irregularly shaped grains in value, but the difference in shape can lead to systematic differences in the opacity index. The largest difference is in the polarization, where spherical grains of large maximum grain sizes produce polarization that is 90$^{\circ}$ offset from that of irregular shaped grains. 
\end{enumerate}

%% IMPORTANT! The old "\acknowledgment" command has be depreciated. It was
%% not robust enough to handle our new dual anonymous review requirements and
%% thus been replaced with the acknowledgment environment. If you try to 
%% compile with \acknowledgment you will get an error print to the screen
%% and in the compiled pdf.
%% 
%% Also note that the akcnowlodgment environment does not support long amounts of text. If you have a lot of people and institutions to acknowledge, do not use this command. Instead, create a new \section{Acknowledgments}.
\section*{acknowledgements}

We thank the anonymous reviewer for the constructive comments that improved the paper.
ZYDL acknowledges support from U.S. National Science Foundation Astronomy and Astrophysics Research Grants \#2307612. 
We are grateful to Maxim Yurkin for helpful advice and for making ADDA publicly available. 
We thank the following people for the help and advice that made this work possible: Erica Behrens, Xi Chen, and Anirudh Prabhu.
We would also like to thank the following people for fruitful discussions: Peter Gao, Matt Lodge, Julia Martikainen, Olga Mu\~noz, and Haifeng Yang. 

%\end{acknowledgements}

%% To help institutions obtain information on the effectiveness of their 
%% telescopes the AAS Journals has created a group of keywords for telescope 
%% facilities.
%
%% Following the acknowledgments section, use the following syntax and the
%% \facility{} or \facilities{} macros to list the keywords of facilities used 
%% in the research for the paper.  Each keyword is check against the master 
%% list during copy editing.  Individual instruments can be provided in 
%% parentheses, after the keyword, but they are not verified.

%\vspace{5mm}
%\facilities{HST(STIS), Swift(XRT and UVOT), AAVSO, CTIO:1.3m,
%CTIO:1.5m,CXO}

%% Similar to \facility{}, there is the optional \software command to allow 
%% authors a place to specify which programs were used during the creation of 
%% the manuscript. Authors should list each code and include either a
%% citation or url to the code inside ()s when available.

\software{ADDA \citep{Yurkin2011}, dsharp\_opac \citep{Birnstiel2018ApJ...869L..45B}, numpy \citep{harris2020array}, scipy \citep{2020SciPy-NMeth}, matplotlib \citep{Hunter:2007}, PyTorch \citep{pytorch2019arXiv191201703P} 
}

%% Appendix material should be preceded with a single \appendix command.
%% There should be a \section command for each appendix. Mark appendix
%% subsections with the same markup you use in the main body of the paper.

%% Each Appendix (indicated with \section) will be lettered A, B, C, etc.
%% The equation counter will reset when it encounters the \appendix
%% command and will number appendix equations (A1), (A2), etc. The
%% Figure and Table counter will not reset.

%% For this sample we use BibTeX plus aasjournals.bst to generate the
%% the bibliography. The sample631.bib file was populated from ADS. To
%% get the citations to show in the compiled file do the following:
%%
%% pdflatex sample631.tex
%% bibtext sample631
%% pdflatex sample631.tex
%% pdflatex sample631.tex

\bibliography{main}{}
\bibliographystyle{aasjournal}

%% This command is needed to show the entire author+affiliation list when
%% the collaboration and author truncation commands are used.  It has to
%% go at the end of the manuscript.
%\allauthors

%% Include this line if you are using the \added, \replaced, \deleted
%% commands to see a summary list of all changes at the end of the article.
%\listofchanges

\end{document}